\documentclass[final,5p,times,twocolumn,authoryear]{elsarticle}

\usepackage{amssymb}
\usepackage{lipsum}
\usepackage{aas_macros}
\usepackage{graphicx}
\usepackage{hyperref}
\usepackage{multirow}
\usepackage{amsmath}
\usepackage{isotope}
\usepackage[dvipsnames]{xcolor}
\newcommand{\msun}{$M_\odot$}
\graphicspath{{figures/}}

\journal{High Energy Astrophysics}

\begin{document}

\begin{frontmatter}

%% Title, authors and addresses

%% use the tnoteref command within \title for footnotes;
%% use the tnotetext command for theassociated footnote;
%% use the fnref command within \author or \affiliation for footnotes;
%% use the fntext command for theassociated footnote;
%% use the corref command within \author for corresponding author footnotes;
%% use the cortext command for theassociated footnote;
%% use the ead command for the email address,
%% and the form \ead[url] for the home page:
%% \title{Title\tnoteref{label1}}
%% \tnotetext[label1]{}
%% \author{Name\corref{cor1}\fnref{label2}}
%% \ead{email address}
%% \ead[url]{home page}
%% \fntext[label2]{}
%% \cortext[cor1]{}
%% \affiliation{organization={},
%%            addressline={}, 
%%            city={},
%%            postcode={}, 
%%            state={},
%%            country={}}
%% \fntext[label3]{}

\title{Propagating Uncertainties from Nuclear Physics to Gamma-rays in Core Collapse Supernovae}

\author[LANL]{Chris L Fryer}

\author[msu]{Hendrik Schatz}

\author[LANL]{Samuel Jones}

\author[ncstate]{Atul Kedia}

\author[ncstate,tunl]{Richard\ Longland}

\author[jena]{Fabio Magistrelli}

\author[uv]{Gerard Nav\'o}

\author[uvic]{Joshua Issa}

\author[asu]{Patrick A Young}

\author[uoy,triumf]{Alison M. Laird}

\author[lsu]{Jeffery C. Blackmon}

\author[tuda,gsi,mpik]{Almudena Arcones}

\author[LANL]{Samuel Cupp}

\author[ncstate]{Carla\ Fr\"ohlich}

\author[uvic]{Falk Herwig}

\author[LANL]{Aimee Hungerford}

\author[umn]{Chen-Qi Li}

\author[ncstate]{G.\ C.\ McLaughlin}

\author[Clemson]{Bradley S. Meyer}

\author[LANL]{Matthew R.\ Mumpower}

\author[umn]{Yong-Zhong Qian}

\affiliation[LANL]{organization={Los Alamos National Laboratory},
            city={Los Alamos},
            postcode={87545}, 
            state={NM},
            country={USA}}
\affiliation[umn]{organization={University of Minnesota},
        	city={Minneapolis},
        	postcode={55455},
        	state={MN},
        	country={USA}}
\affiliation[ncstate]{organization={North Carolina State University},
            city={Raleigh},
            postcode={27695},
            state={NC},
        	country={USA}}
\affiliation[tunl]{organization={Triangle Universities Nuclear Laboratory},
            city={Durham},
            postcode={27708},
            state={NC},
            country={USA}
}
\affiliation[uv]{organization={Universitat de Valencia},
            city={Burjassot},
            postcode={46100},
            country={Spain}
}
\affiliation[jena]{organization={Theoretisch-Physikalisches Institut, Friedrich-Schiller-Universität Jena},
            city={Jena},
            postcode={07743},
            country={Germany}}

\affiliation[uvic]{organization={University of Victoria},
            city={Victoria},
            postcode={V8P 5C2},
            country={Canada}}

\affiliation[msu]{organization={Michigan State University},
            city={East Lansing},
            postcode={48864},
            state={MI},
        	country={USA}}
\affiliation[uoy]{organization={School of Physics, Engineering and Technology, University of York},
            city={York},
            postcode={YO10 5DD},
            country={UK}}

\affiliation[lsu]{organization={Louisiana State University},
            city={Baton Rouge},
            postcode={70803},
            state={LA},
            country={USA}}

\affiliation[tuda]{organization={Technische Universität Darmstadt},
        	city={Darmstadt},
        	postcode={64289},
        	country={Germany}}
\affiliation[gsi]{organization={GSI Helmholtzzentrum für Schwerionenforschung GmbH},
        	city={Darmstadt},
        	postcode={64291},
        	country={Germany}}
\affiliation[mpik]{organization={Max-Planck-Institut für Kernphysik},
        	city={Heidelberg},
        	postcode={69117},
        	country={Germany}} 
\affiliation[asu]{organization={Arizona State University},
            city={Tempe},
            postcode={85287},
            state={AZ},
            country={USA}}
\affiliation[Clemson]{organization={Clemson University},
            city={Clemson},
            postcode={29634},
            state={SC},
            country={USA}}
\begin{abstract}

Nuclear yields are powerful probes of supernova explosions, their engines and their progenitors.  In addition, as we improve our understanding of these explosions, we can use nuclear yields to probe dense matter and neutrino physics, both of which play a critical role in the central supernova engine.  Especially with upcoming gamma-ray detectors that can directly detect radioactive isotopes out to increasing distances from gamma-rays emitted during their decay, nuclear yields have the potential to provide some of the most direct probes of supernova engines and stellar burning.  To utilize these probes, we must understand and limit the uncertainties in their production.  Uncertainties in the nuclear physics can be minimized by combining both laboratory experiments and nuclear theory.  Similarly, astrophysical uncertainties caused by simplified explosion trajectories can be minimized by higher-fidelity stellar-evolution and supernova-engine models.  This paper reviews the physics and astrophysics uncertainties in modeling nucleosynthetic yields, identifying the key areas of study needed to maximize the potential of supernova yields as probes of astrophysical transients and dense-matter physics.

\end{abstract}

%%Graphical abstract
%\begin{graphicalabstract}
%\includegraphics{grabs}
%\end{graphicalabstract}

%%Research highlights
%\begin{highlights}
%\item Research highlight 1
%\item Research highlight 2
%\end{highlights}

\begin{keyword}
%% keywords here, in the form: keyword \sep keyword, up to a maximum of 6 keywords
Nuclear Astrophysics: nucleosynthesis in novae and supernovae, 26.30.-k \sep Supernovae, 97.60.Bw \sep Supernovae, nucleosynthesis in, 26.30.-k 

%% PACS codes here, in the form: \PACS code \sep code

%% MSC codes here, in the form: \MSC code \sep code
%% or \MSC[2008] code \sep code (2000 is the default)

\end{keyword}

\end{frontmatter}

%\tableofcontents

%% \linenumbers

%% main text

\section{Introduction}
\label{sec:intro}

Massive stars and core-collapse supernovae (SNe) are major sources of elements heavier than carbon.  The elements produced in a given explosion depend on both the progenitor and the nature of the explosion.  Our understanding of massive star evolution and the explosive engine is still limited by uncertainties in the astrophysical modeling.  Both jet-driven~\citep[e.g.][]{2014MNRAS.439.4011G,2018MNRAS.477.2366H,2023ApJ...948...80L,2024ApJ...969..163W} and convection-enhanced neutrino-driven~\citep[e.g.][]{1994ApJ...435..339H,2007ApJ...659.1438F,2014ApJ...786...83T,2015ApJ...807L..31L,2018SSRv..214...33B,navo+2023,sieverding2023ti44} engines are being actively studied and, at this time, it is unknown what fraction of observed core-collapse supernovae are powered by each engine.  Similarly, recent studies of stellar convection and its effect on shell-burning and the merger of burning layers can dramatically alter the structure of massive stars~\citep{2005A&A...443..643Y,2019ApJ...882...18A,2023MNRAS.525.1601H,2024MNRAS.533..687R}.  In turn, these modified structures affect both the nature of the supernova explosion and its production of heavy elements.

The observation of the decay products from radioactive nuclei provides one of the most direct observations of these astrophysical models.  The early rise in the $\gamma$-rays from $^{56}$Ni decay from SN 1987A was an early indication that the SN engine had strong asymmetries.  $^{56}$Ni is produced in the innermost ejecta and, as such, any decay photons would be trapped in the ejecta, down-scattering to low energies before escaping~\citep{1988ApJ...329..820P}.  Instead, for SN 1987A, the $\gamma$-rays from $^{56}$Ni were observed early, arguing that this $^{56}$Ni, intially produced in the central engine, somehow mixed out into the outer ejecta layers~\citep{1988ApJ...329..820P}.  This extensive mixing led to the development of the currently-favored convection-enhanced supernova engine paradigm~\citep{1994ApJ...435..339H}.  We will refer to this as the convective engine in this paper.  Further observations of the Doppler broadening of the $\gamma$-ray lines indicated a strong component moving away from our line-of-sight~\citep{2003ApJ...594..390H,2005ApJ...635..487H}, providing further support for this convection-enhanced engine for SN 1987A. Finally, maps of $^{44}$Ti (also produced in this innermost ejecta) in the Cassiopeia A remnant~\citep{2014Natur.506..339G,2017ApJ...834...19G} provided strong validation of the convective engine for normal core-collapse SNe.  

As observations of SNe and SN remnants improve, we can further constrain our models.  To use observations of nucleosynthetic yields, we must identify and constrain the uncertainties in their production.  The uncertainties span a wide range of properties of SNe including the modeling of progenitor evolution (e.g. stellar mass loss, role of rotation, shell burning and mixing), supernova engines (e.g. dense nuclear matter and neutrino physics, neutrino transport, turbulence and fluid-dynamics models) and the trajectory evolution (e.g. including the effects of long-lived engine activity, shock capturing capabilities and multi-dimensional effects of shock propagation).  

Even with high-fidelity astrophysics calculations allowing astronomers to focus on the supernova engine properties and physics (e.g. jet-driven versus convection-enhanced neutrino-driven, neutrino and dense-matter physics), uncertainties in the nuclear cross sections and rates can limit the use of nucleoynsthetic yields as a probe of our understanding of supernovae.  A coupled theory and experimental program is required to reduce these uncertainties.

To tie to many observations, additional physics is required in radiation transport and radiation-hydrodynamics, out-of-equilibrium physics, atomic physics (and the inclusion of out-of-equilibrium effects on the atomic level states), and the effects of a complex circumstellar medium (including dust obscuration and shock heating).  One of the most direct probes is the observation of decay lines from radioactive isotopes.  Although most of the current efforts have focused on $^{56}$Ni and $^{44}$Ti production, a broad number of radioactive isotopes are promising (Table~\ref{tab:decay} - data is from the Nuclear Data Services website:  \url{https://www-nds.iaea.org/relnsd/vcharthtml/VChartHTML.html}).  These isotopes allow us to probe both stellar evolution (stellar burning layers and mixing) and the supernova engine.

\begin{table}
  \centering
  \scriptsize
  \caption{List of the radioactive nuclei and their decay lines used in this study.}
  \begin{tabular}{l|cc}
    \hline
    Isotope & E$_{\gamma-ray}$ & Decay Percentage \\
    \hline
    \hline
    & & \\
    \isotope[56]{Ni} $\rightarrow$ \isotope[56]{Co} & 6.915 keV & 10\% \\
    t$_{1/2}=$6.075d & 6.93 keV & 19.7\% \\
    & 158.38 keV & 98.8\% \\
    & 269.50 keV & 36.5\% \\
    & 480.44 keV & 36.5\% \\
    & 749.95 keV & 49.5\% \\
    & 811.85 keV & 86.0\% \\
    & 1561.80 keV & 14.0\% \\
    \hline
    & & \\
    \isotope[56]{Co} $\rightarrow$ \isotope[56]{Fe} & 846.8 keV & 99.9\% \\
    t$_{1/2}=$77.24d & 1037.8 keV & 14.1\% \\
    & 1238.29 keV & 66.5\% \\
    & 1771.36 keV & 15.4\% \\
    & 2598.50 keV & 17.0\% \\
    \hline
    \hline
    & & \\
    \isotope[47]{Ca} $\rightarrow$ \isotope[47]{Sc} & 1297.1 keV & 67.0\% \\
    t$_{1/2}=$4.536d &  &  \\
    \hline
    & & \\
    \isotope[47]{Sc} $\rightarrow$ \isotope[47]{Ti} & 159.4 keV & 68.3\% \\
    t$_{1/2}=$3.3492d & &  \\
    \hline
    \hline
    & & \\
    \isotope[43]{K} $\rightarrow$ \isotope[43]{Ca} & 372.8 keV & 86.8\% \\
    t$_{1/2}=$22.3 h & 396.9 keV & 11.9\% \\
    & 593.4 keV & 11.3\% \\
    & 617.5 keV & 79.2\% \\
    \hline
    \hline
    & & \\
    \isotope[48]{Cr} $\rightarrow$ \isotope[48]{V} & 4.95 keV & 6.4/12.9\% \\
    t$_{1/2}=$21.56 h & 112.31 keV & 96.0\% \\
    & 308.2 keV & 100\% \\
    \hline
    & & \\
    \isotope[48]{V} $\rightarrow$ \isotope[48]{Ti} & 983.5 keV & 100\% \\
    t$_{1/2}=$15.97 d & 1312.1 keV & 98.2\% \\
    \hline
    \hline
    & & \\
    \isotope[51]{Cr} $\rightarrow$ \isotope[51]{V} & 4.95 keV & 6.5/12.9\% \\
    t$_{1/2}=$27.704 d &  &  \\
    \hline
    \hline
    & & \\
    \isotope[52]{Mn} $\rightarrow$ \isotope[52]{Cr} & 744.2 keV & 90\% \\
    t$_{1/2}=$21.1 min & 935.5 keV & 94.5\% \\
    & 1434.1 keV & 100.0\% \\
    \hline
    \hline
    & & \\
    \isotope[59]{Fe} $\rightarrow$ \isotope[59]{Co} & 1099.2 keV & 56.5\% \\
    t$_{1/2}=$44.49 d & 1291.6 keV & 43.2\% \\
    \hline
    \hline
    & & \\
    \isotope[57]{Ni} $\rightarrow$ \isotope[57]{Co} &  127.2 keV & 16.7\% \\
    t$_{1/2}=$35.6 h & 1377.6 keV & 81.7\% \\
    & 1919.52 keV & 12.3\% \\
    \hline
    & & \\
    \isotope[57]{Co} $\rightarrow$ \isotope[57]{Fe} & 6.391 keV & 16.6\% \\
    t$_{1/2}=$271.74 d & 6.404 keV & 32.9\% \\
    & 122.1 keV & 85.6\% \\
    & 136.5 keV & 10.7\% \\
    \hline
    \hline
    & & \\
    \isotope[60]{Fe} $\rightarrow$ \isotope[60]{Co} &  6.9 keV & 9.1/18.0\% \\
    t$_{1/2}= 2.62 \times 10^6$y & 1377.6 keV & 81.7\% \\
    & 1919.52 keV & 12.3\% \\
    \hline
    \hline
    & & \\
    \isotope[26]{Al} $\rightarrow$ \isotope[26]{Mg} & 1807.7 keV & 99.8\% \\
    t$_{1/2}=7.17\times10^5$y &  \\
    \hline
    \hline
  \end{tabular}
  \label{tab:decay}
\end{table}

In this paper, we review both the astrophysical (Section~\ref{sec:astro}) and nuclear physics (Section~\ref{sec:nuclear}) uncertainties, studying the effects these uncertainties have on nuclear yields and radioactive isotopes in particular.  In Section~\ref{sec:examples}, we include a few focused examples for the yields in the innermost ejecta (direct probes of the supernova engine) and shell-burning trajectories (probing stellar evolution).  We summarize our results in Section~\ref{sec:conclusions}.

\section{Astrophysical Uncertainties}
\label{sec:astro}

For core-collapse supernovae, uncertainties in our understanding of stellar progenitors, the characteristics of the explosive engine, the propagation of the SN shock through the star and the calculation of the observables (in our case, $\gamma$-rays) all contribute to the uncertainties in the nuclear yields and their effect on the observations.  Here we review those uncertainties.

\subsection{Progenitors}

Many astrophysical fields rely on an accurate understanding of stars and stellar evolution.  Unfortunately, stellar evolution is one of the most challenging of astrophysical problems numerically.  Numerically resolved simulations of stellar evolution in either space and time is simply not possible with current and future computing.  Instead, stellar modellers use a set of prescriptions that attempt to capture the physics in stellar evolution.  For many years, the simplified stellar models were sufficient to match the myriad of observations using them from stellar observations to supernova explosions.  However, as broader and more-accurate observations have placed increasingly precise constraints on the models, we are beginning to understand the limitations of these prescriptions.  Here we review the uncertainties in our stellar models and how they might affect the yields.

The uncertainties in our stellar models include:
\begin{itemize}
    \item {\bf Stellar Mixing:}  Most stellar evolution codes use Mixing Length Theory~\citep{1958ZA.....46..108B} with an increasing number of additions such as ``semi-convection'' and ``over-shooting'' to try to capture the physics of stellar convection. It should be noted that add-ons to mixing length theory are not predictive. As we shall discuss in more detail below, uncertainties in this method lead to shell mergers that can drastically alter the pre-collapse structure of the star, dramatically altering the fate of the collapse and the yields. Hydrodynamic simulations of stellar convection indicate that these shells mergers do happen.
    \item {\bf Stellar Burning:}  Although not the focus of this paper, uncertainties in the nuclear physics, e.g. $^{12}{\rm C}(\alpha,\gamma)^{16}{\rm O}$, $^{22}{\rm Ne}(\alpha,n)^{25}{\rm Mg}$, can also dramatically change the structure of the star as well as the neutron content that then can alter the explosive yields.  In addition, the implementation of the nuclear reactions, especially for complex reaction processes, like Silicon burning, can lead to  large uncertainties in the final structure of the star~\citep{2018MNRAS.480..538R}.
    \item {\bf Radiation Transport} is typically modeled using a single-temperature diffusion approximation.  As we shall discuss below, this can miscalculate the energy transport through radiation by up to 20\% in a star.
    \item {\bf Stellar Rotation} can have a dramatic effect on many aspects of stellar evolution~\citep{2000A&A...361..101M}, including leading to extensive shell mergers in stellar mixing that dramatically alters the structure of the star~\citep{2005A&A...443..643Y}.
    \item {\bf Stellar Mass Loss} can also alter the structure of the star and its core temperature, %altering 
    thereby affecting the yields~\citep{2025A&A...699A..71H}.  However, as we shall discuss below, stellar mass loss (and even the nature of the stellar envelope) is very poorly understood. Interestingly, neutrinos emitted subsequent to the core collapse reach the envelope way before the supernova shock, and gamma rays from neutrino capture on protons and positron annihilation with electrons following absorption of electron antineutrinos by protons may provide useful probes of the density structure of the stellar envelope~\citep{Lu+2007,Lunardini+2024}.
\end{itemize}

The long timescales in stellar evolution make full hydrodynamic calculations of stellar mixing intractable.  In most stellar evolution codes, stellar mixing is incorporated using mixing-length theory~\citep{1958ZA.....46..108B}.  However, this method is approximate at best~\citep{2019ApJ...882...18A} and more advanced solutions exist, e.g. Reynolds Averaged Navier Stokes~\citep{2015ApJ...809...30A}.  A growing number of multi-dimensional simulations exist to understand this physics~\citep{2023MNRAS.525.1601H,2024MNRAS.533..687R}, but these calculations are limited to snapshots in stellar models.  We discuss the errors associated from mixing for a specific example of O/C shell burning in Section~\ref{sec:stellarmixing}.

\begin{figure}[h]
  \centering
  \includegraphics[width=0.48\textwidth]{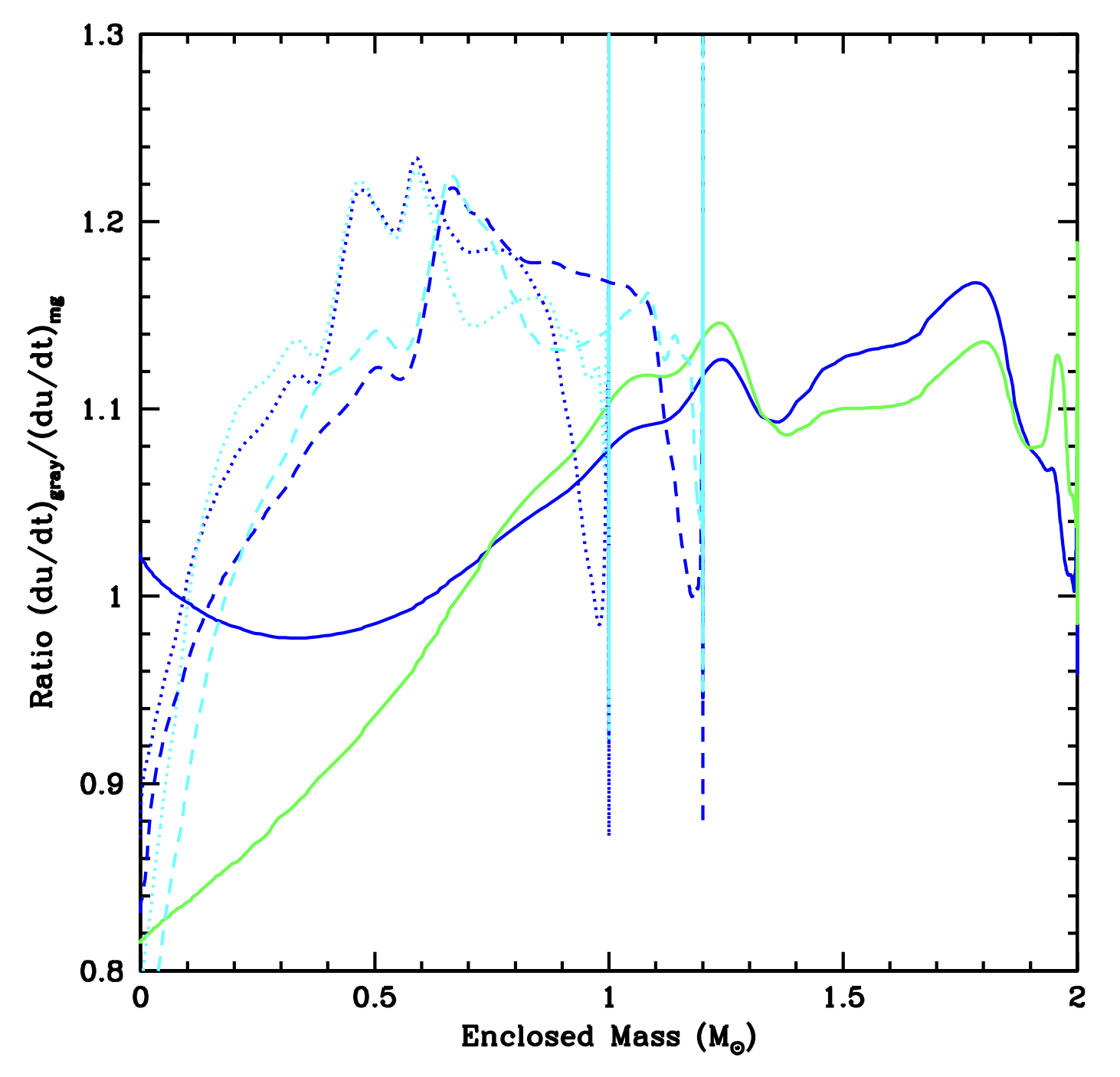}
    \caption{Comparison of the energy transport between a Rosseland gray (single group) opacity and a 100 (multi-)group opacity for a 2 (solid), 1.2 (dashed), 1.0 (dotted) $M_\odot$ star.  The different colors correspond to different compositions for the star.}
    \label{fig:star_tran}
\end{figure}

Energy generated in the core or shell burning regions of a star propagates out through the star either through stellar convection, conduction or radiation transport. The efficacy of conduction or radiation transport affect the temperature structure of the star and will, ultimately, dictate where convection occurs in the star.  Most stellar evolution codes use 1-T diffusion to model the radiation transport.  The diffusion approximation in radiation transport can be derived by integrating over angle the Boltzmann transport equation, a.k.a. a first moment solution~\citep[for a review, see][]{2004rahy.book.....C}.  This effectively assumes a specific angular distribution for the photons (a.k.a. a Lambertian).  1-T refers to the fact that the radiation is assumed to be a blackbody and the matter is a Maxwellian, both described by a single temperature.  In this 1-T approximation, the opacity is averaged over all energies and is typically calculated using a Rosseland-averaged opacity, also called gray transport~\citep{2004rahy.book.....C}.  Although the high stellar densities would justify the diffusion approximation, the approximation of a {\bf single} Rosseland-mean opacity can cause up to 20\% errors in the energy transport (Figure~\ref{fig:star_tran}).  The gray transport approximation overestimates the energy transport leading to a different stellar structure.  If 10-20\% effects are important, we must improve our transport schemes in our stars.

Another major challenge in stellar evolution is the modeling of rotation including: 1) understanding how viscous forces couple the rotation in different burning layers, 2) determining the effect rotation has on mixing and 3) calculating the angular momentum loss through winds.  As a star evolves, it develops a series of composition layers from the ashes of different core-burning stages.  If limited to microphysical viscous forces, the different shells would not be strongly coupled, allowing the core to continue to spin rapidly even as the star expands in a giant phase.  However, it has been shown that magnetic fields can be generated by the differential rotation~\citep{2002A&A...381..923S} and this ``magnetic viscosity'' can strongly couple the shells, producing slowly rotating cores.  This can have profound effects on the supernova engine.  In addition, the differential rotation can induce mixing between shells, causing these shells to merge~\citep{2005A&A...443..643Y} adding to our uncertainties in stellar mixing.  In modern calculations, stellar evolution codes input this coupling with a free parameter for the growth of the magnetic fields, but analyses of the conditions in stars show that the physics is much more complex than this~\citep{2003A&A...411..543M}.  Understanding this physics is critical to understanding the yields in supernovae.

Although a basic understanding of stellar winds has existed for over 50 years~\citep{1975ApJ...195..157C}, the strength of this wind continues to evolve~\citep{2000ARA&A..38..613K} with observations, more than theory, guiding our understanding.  Rotation can also alter the wind mass-loss~\citep{1984ApJ...284..337O,2000ARA&A..38..143M, 2008A&ARv..16..209P,2018Natur.561..498J,2019MNRAS.485..988O}.  Mass loss from winds can dramatically alter the fate of a star~\citep{2003ApJ...591..288H}.  It is also becoming increasingly evident that the simple line-driven wind picture does not capture all of the possible mass-loss mechanisms.  Explosive shell burning, opacity-driven instabilities and pressure waves can all produce bursts of mass ejection in a star~\citep{2006ApJ...647.1269F,2014ApJ...792L...3H,2016MNRAS.458.1214Q}.  

The radii of massive stars is also difficult to determine.  Stellar models assume a pressure for the outer boundary condition.  The choice of this outer boundary condition can produce a range of results and these differences can vary the fate of massive star binaries~\citep{1999ApJ...526..152F}.  For massive stars, both red giants and Wolf-Rayet stars, the boundary between stellar boundary and stellar outflow is difficult to determine~\citep{2012A&A...538A..40G,2022ApJ...929..156G}.

The uncertainties from all of these physical approximations used to model massive stars alter the stellar structure at collapse.  The structure of the innermost $\sim 3\,M_\odot$ dictates the nature of the supernova explosion and differences can mean the difference between a strong explosion and the collapse to a black hole.  Even small differences will change the strength of the supernova shock.  It is through the structure of the rest of the star that the supernova shock propagates to produce explosive yields.  With systematic studies of the yields, we may be able to guide improved stellar models.  But, to do so, we must understand the nature of the interaction of the supernova shock wave with the star and this requires understanding the evolution of the ejecta, or explosive trajectories. 

\subsubsection{Mixing Uncertainties for O-C Merger Nucleosynthesis}
\label{sec:stellarmixing}

\begin{figure}[!ht]
  \centering
  \includegraphics[width=0.5\textwidth]{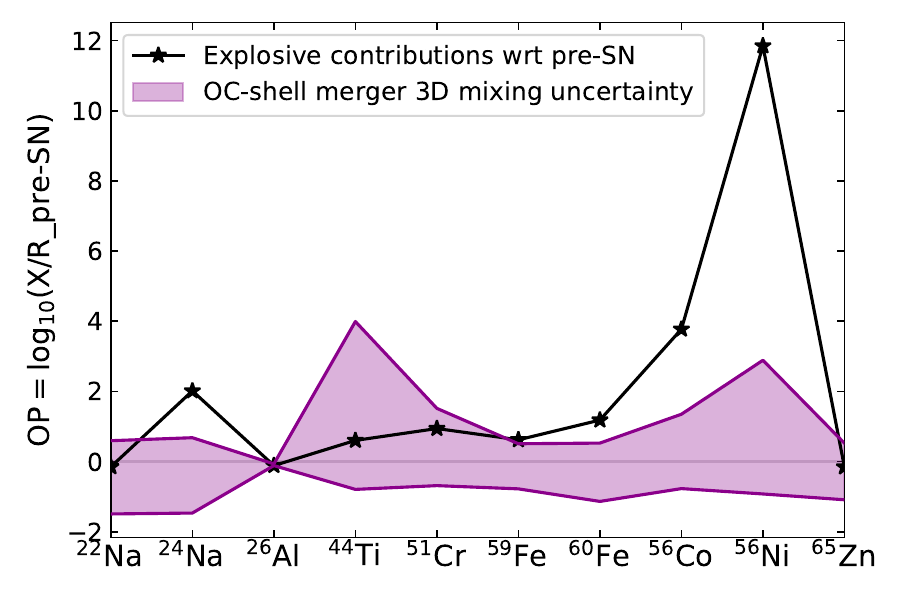}
    \caption{Comparison of the OC-shell merger mixing uncertainties \citep{issaImpact3DMacro2025,issa3DMacroPhysics2025} and explosive nucleosynthesis \citep{2018MNRAS.480..538R} to pre-supernova nucleosynthesis  \citep{2018MNRAS.480..538R} for radioactive isotopes.}
    \label{fig:ocmerger_comp}
\end{figure}

We briefly discussed stellar mixing above.  Here we present a deep dive into the effect of these mixing uncertainties on the O-C shell merger.  O-C shell mergers in massive stars are events that alter the progenitor structure hours to minutes before core-collapse when the O and C shells merge into a single convective zone. 1D stellar models cannot replicate the fundamentally 3D nature of the macro physics in convective O shell burning since this involves non-radial spherical asymmetric instabilities and plume-driven mixing \citep{jonesIdealizedHydrodynamicSimulations2017, andrassy3DHydrodynamicSimulations2020, rizzutiShellMergersLate2024a}.
Because the O shell is a convective-reactive environment where timescales for mixing and burning are comparable, changes to the mixing profile can have significant impact on the nucleosynthesis.
Exploratory research into the impact of 3D macro physics on 1D nucleosynthesis has shown that uncertainties due to mixing are large for both light and heavy isotopes synthesized in the O shell during a merger \citep{issaImpact3DMacro2025, issa3DMacroPhysics2025}.
Production during the O-C shell merger can dominate both pre-explosive and explosive nucleosynthesis for light and heavy isotopes, including $^{44}\mathrm{Ti}$ \citep{robertiGprocessNucleosynthesisCorecollapse2024,robertiOccurrenceImpactCarbonoxygen2025}.
Using post-processed models that implement 3D macro physical mixing prescriptions \cite{issaImpact3DMacro2025, issa3DMacroPhysics2025} into the 1D $15~\mathrm{M_\odot}$ $Z=0.02$ NuGrid model \citep{2018MNRAS.480..538R}, Figure \ref{fig:ocmerger_comp} demonstrates that radioactive isotopes abundances right before explosion, including $^{44}\mathrm{Ti}$, can be changed by orders of magnitude.

\begin{figure}[!ht]
  \centering
  \includegraphics[width=0.5\textwidth]{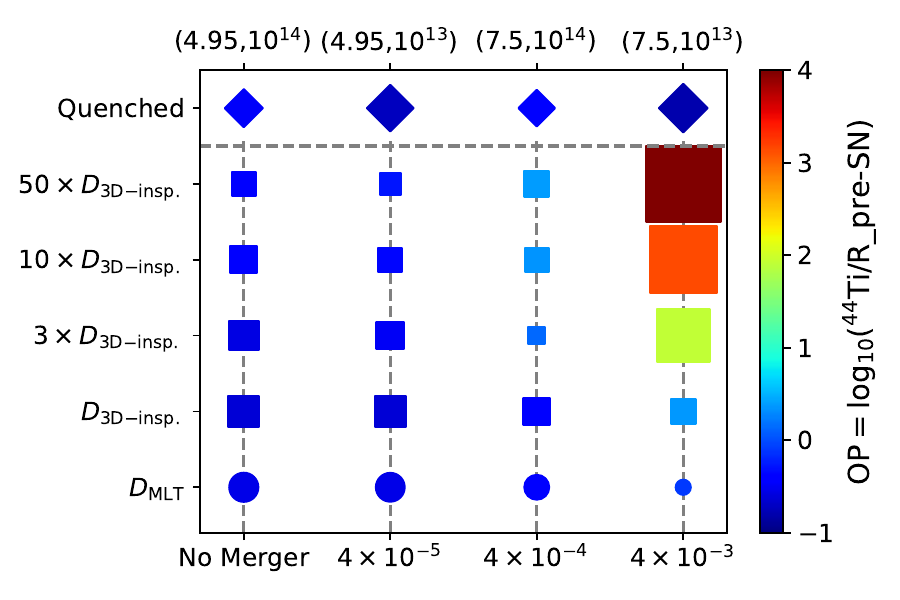}
    \caption{Overproduction of $^{44}\mathrm{Ti}$ for the mixing cases described in \cite{issaImpact3DMacro2025,issa3DMacroPhysics2025}. The lower x-axis is the rate of ingesting C-shell material into the O shell in $\mathrm{M_\odot}\mathrm{s}^{-1}$ for the 1D mixing length theory (circles) and 3D-inspired with mixing profile downturn (squares) cases. The upper x-axis denotes quenched (diamonds) cases and are labelled by (center of the mixing efficiency dip in Megameters, extent of the dip in $\mathrm{cm}^{2}\mathrm{s}^{-1}$). Size indicates distance from $\mathrm{OP}=0$ and color indicates magnitude. Median $\mathrm{OP}=-0.4$ and the explosive production of $^{44}\mathrm{Ti}$ \citep{2018MNRAS.480..538R} has an $\mathrm{OP}=0.6$}
    \label{fig:ocmerger_ti44}
\end{figure}

The OC-shell merger 3D mixing uncertainty for these isotopes is $2.2~\mathrm{dex}$ on average and $^{44}\mathrm{Ti}$ features the largest uncertainty of $4.8~\mathrm{dex}$.
These pre-explosive merger uncertainties are comparable to or larger than the explosive results for $^{22}\mathrm{Na}$, $^{26}\mathrm{Al}$, $^{44}\mathrm{Ti}$, $^{51}\mathrm{Cr}$, $^{59}\mathrm{Fe}$, and $^{65}\mathrm{Zn}$. 
Figure \ref{fig:ocmerger_ti44} shows that not all merger scenarios equally impact the production of $^{44}\mathrm{Ti}$.
Models with ingestion rates of a full merger that implement boosted convective velocities with a mixing profile downturn show the largest enhancement, though a majority of models have less $^{44}\mathrm{Ti}$ than the pre-supernova results of \cite{2018MNRAS.480..538R}.
20\% of massive stars models feature an O-C shell mergers \citep{robertiOccurrenceImpactCarbonoxygen2025}, and although different stars will not have the same mixing conditions during their mergers, these results show that the inclusion of realistic 3D scenarios can dramatically change production of radioactive isotopes in the O-C shell before the onset of a core-collapse supernova.

\subsection{Explosive Trajectories}

The nucleosynthetic yields can be determined by the time evolution of the density, $\rho(t)$, and temperature, $T(t)$, profiles.  The simple picture for explosive yields from supernovae assumes that, once the shock is launched, the ejecta quickly accelerates and then expands ballistically.  During acceleration, the ejecta evolution is typically given by~\citep{1964ApJ...139..909H}:
\begin{equation}
    d\rho/dt = - \rho/\tau
\end{equation}
where $\tau$ is the acceleration timescale.  If we assume the ejecta is adiabatic for a radiation dominated gas ($T^3/\rho \propto S$ where the entropy, $S$, is constant in adiabatic expansion), the temperature is then given by:
\begin{equation}
   dT/dt = - T/(3 \tau). \label{eq:adiabatic1}
\end{equation}
These yield an exponential decay in the density and temperature:
\begin{equation}
    \rho = - \rho_0 e^{-t/\tau},
\end{equation}
\begin{equation}
    T = - T_0 e^{-t/(3 \tau)}.
\end{equation}
where $\rho_0$ and $T_0$ are the initial density and temperature, respectively, when the shock is launched. After the material becomes ballistic (end of acceleration phase), the density decreases with increasing volume, a power law shock is often used~\cite{2010ApJS..191...66M}:
\begin{equation}
    \rho(t) = \rho_0/(t/t_{exp} +1)^3 \label{eq:adiabatic2}
\end{equation}
A number of studies have focused on these trajectories~\citep{2010ApJS..191...66M} and, as we shall see for some of our applications (Section~\ref{sec:shell}), these trajectories provide a good first understanding of nucleosynthetic yields in core-collapse supernovae.

But deviations from these simple trajectories exist that can alter the nucleosynthetic yields.  Although some effects can occur in our 1-dimensional trajectories, many are only seen in multi-dimensional models.  We briefly review some of these effects here.

\begin{figure}[!ht]
  \centering
  \includegraphics[width=0.5\textwidth]{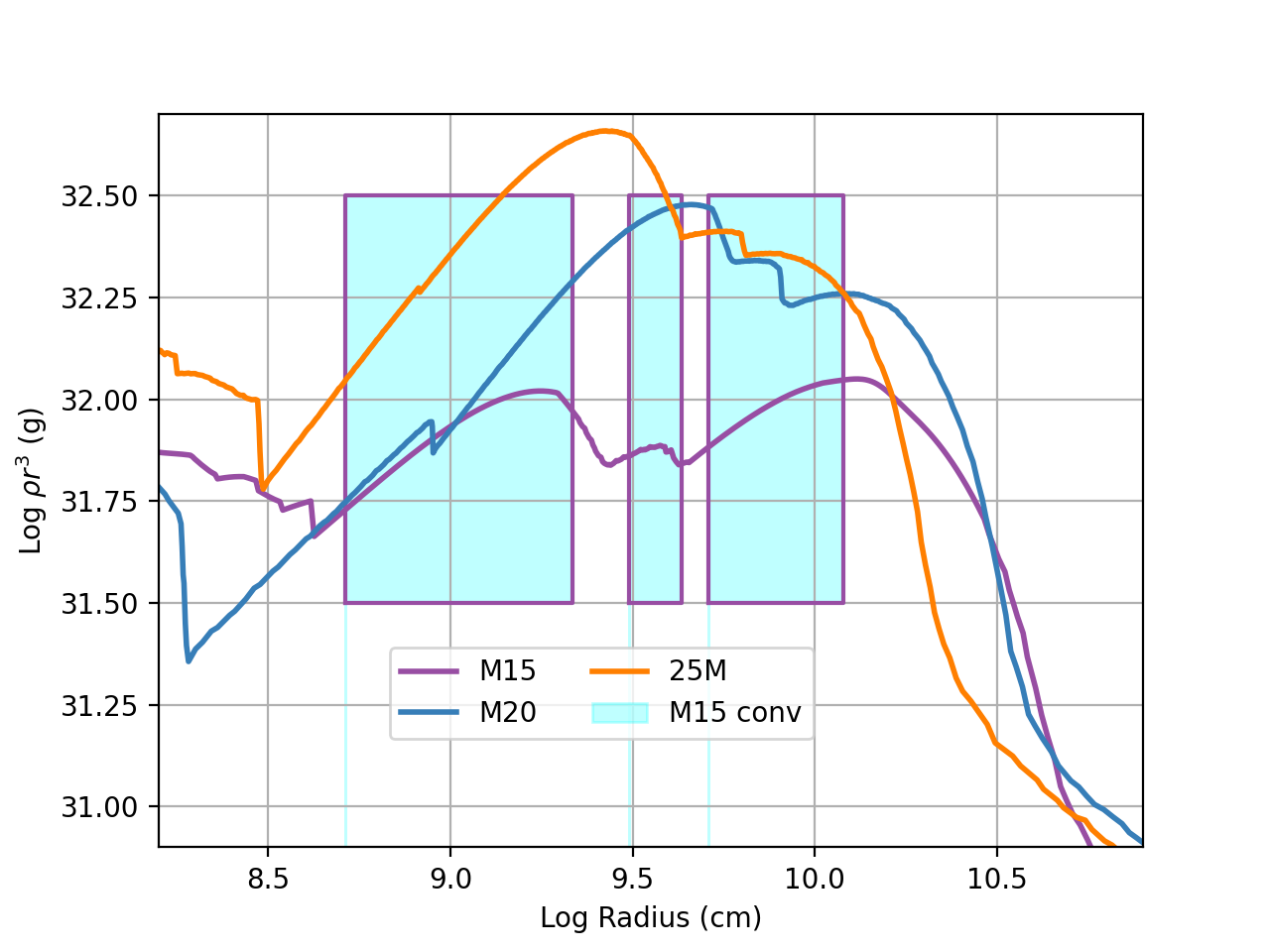}
    \caption{Log of density times radius cubed for 3 different stellar models as a function of radius in a star for 3 different stellar models from ~\cite{2002NewAR..46..463H}:  15, 20, and 25\,M$_{\odot}$.  If this value is increasing, $\alpha < 3$ and the shock will decelerate.  Conversely, if it is decreasing, the shock will accelerate.  $\rho r^3$ tends to increase in convective regions, decreasing in radiative regions of the star (the shaded regions correspond to convective regions in the 15\,M$_\odot$ star.}
    \label{fig:dec-star}
\end{figure}

{\it Shock deceleration:}  
Exponential trajectories capture an acceleration of the ejecta and power-law trajectories capture ballistic trajectory.  Supernova ejecta tends to decelerate with time.  As the supernova shock moves out through the star, its velocity can both increase or decrease based on the density profile of the star.  This propagation is well-described by the Taylor–von Neumann–Sedov similarity solution~\citep{1959sdmm.book.....S}:
\begin{equation}
 v_{\rm shock} \propto v_0 (t/t_0)^{(\alpha-3)/(5-\alpha)}
\end{equation}
where $\alpha$ denotes the density gradient ($\rho \propto r^{-\alpha}$), $v_{0}$ is the initial velocity when the shock starts decelerating, and $t_0$ is the time at the onset of the deceleration.  If $\alpha>3$, the shock accelerates and if $\alpha<3$, it decelerates.  We can understand the expected regions for acceleration and deceleration by plotting the quantity $\rho r^3$ which increases when $\alpha<3$ and decreases when $\alpha>3$.  Figure~\ref{fig:dec-star} shows the value $\rho r^3$ for 3 stellar models, 15, 20, 25\,M$_{\odot}$.  Deceleration will alter the trajectory and can also cause shocks that change the entropy.  

{\it Continuous Engine:}  In 1-dimension, the launch of the shock typically means the end of the engine.  Aside from any shock deceleration effects, the ejecta will expand ballistically and adiabatically cool.  Without material near the neutron star surface, the engine effectively turns off.  But in multi-dimensional models, continued accretion onto the newly formed neutron star will drive outflows long after the launch of the shock.  This continued ejecta can catch the initial ejecta, causing it to compress, raising its temperature and density.  This rise is seen in many of the trajectories of the innermost ejecta from a core-collapse supernova. Due to the fundamentally 3D nature of the flows, the timing and degree (and even number) of temperature increases results in substantially varying trajectories for material under similar progenitor conditions in different angular directions~\citep{sieverding2023ti44}.

{\it Auxillary Shocks and Compression:}  Multi-dimensional studies also highlight additional effects that can alter the density/temperature evolution of the ejecta.  With the aspherical explosions expected in both convective and jet-driven engines, we expect a range of non-radial shocks and/or compression.  If these effects can reheat the material above $10^9 \, {\rm K}$ (as is the case for explosive shell burning), auxillary shocks can play an important role in the final yields.

Significant non-radial asymmetries can also be imposed on the explosion initial conditions by convection in the burning shells of the progenitor. Low mode (in analogy with spherical harmonics) distortions in isosurfaces may have $\delta$r of 10's of percent from the average r of the surface, amplified during the collapse, and flows can have Mach numbers of several tenths. These effects are smaller than engine-driven re-heating.

For this paper, we have approximated this reheating using a simple parametric form  
\begin{equation}
  \label{eq:bradsTrajectory}
    T_9(t) = a * \exp(-t/\tau_1) + b * t^n * (1 - \tanh((t - t_0)/\tau_2)
\end{equation}
Here, the first term describes the initial drop in temperature while the second term describes the reheating and subsequent temperature decline.  $a$ is the starting temperature, and $\tau_1$ sets the timescale for the initial temperature decline.  $b$, $n$, and $t_0$ shape the reheating, and $\tau_2$ shapes the subsequent temperature decline after the reheating.  A Jupyter notebook is available to generate appropriate files with this trajectory parameterization \cite{Meyer_trajectory_maker_2025}.

{\it Turbulence:}  Mixing of material from different regions of the progenitor star during the explosion could generate novel nucleosynthesis yields by mixing together reactants that aren't normally mixed together. This would require microscopic mixing on short enough timescales for temperatures to remain high enough for burning to take place. This is most likely to take place in turbulent Kelvin-Helmholtz (KH) instabilities at interfaces between two materials. High mode asymmetries primarily driven by Rayleigh-Taylor/Richtmeyer-Meshkov (RT) instabilities at density/entropy boundaries would create the highest surface to volume ratio opportunities for mixing and the highest chance for structures to be fully mixed.

In order to evaluate whether mixing during explosive burning may have a substantial impact, we employ an analytic estimate of RT and KH growth timescales for the Si/O and CO/He shell interfaces. The growth time for RT instabilities in a non-constant entropy medium is derived from the Br\"{u}nt-V\"{a}is\"{a}l\"{a} frequency N. In the limit of radiation-dominated gas,
\begin{equation}
    N^2 \approx \frac{1}{S}\frac{\Delta S}{\Delta r}\frac{\Delta v}{\Delta t}
\end{equation}
where $S$ is the radiation entropy, $\frac{\Delta v}{\Delta t}$ is the acceleration in the direction opposite the entropy gradient, and $\Delta r$ is the width of the interface. For the Si/O shell interface the RT timescale is $\sim 0.8s$, and $\sim 30s$ for the CO/He shell interface.  

The RT instabilities provide large-spatial-scale advection to move material with different starting compositions into proximity. Microscopic mixing allowing burning of mixed material arises from KH instabilities. The characteristic timescale for growth of KH modes in a compressible fluid with moderate or weak magnetic fields depend upon the Mach number $M$ of the shear flow and the wave number of the mode. The velocity of RT growth generates $M \sim 0.1$ or higher. The shortest growth timescale for such Mach numbers is $>$5s for a normalized wave number of $\alpha \sim$ a few tenths \citep{1969Tell...21..167B,1982PhRvL..49..779M}. For this very simplified estimate, we might expect a few percent of the volume of an RT finger originating near the O/Si interface to mix over the duration of explosive burning. Given the extremely approximate treatment, we may surmise that mixing during burning is not a dominant process for abundant isotopes, but is worth exploring computationally for trace isotopes, e.g. very short-lived radioisotopes.

\begin{figure}[!ht]
  \centering
  \includegraphics[width=0.5\textwidth]{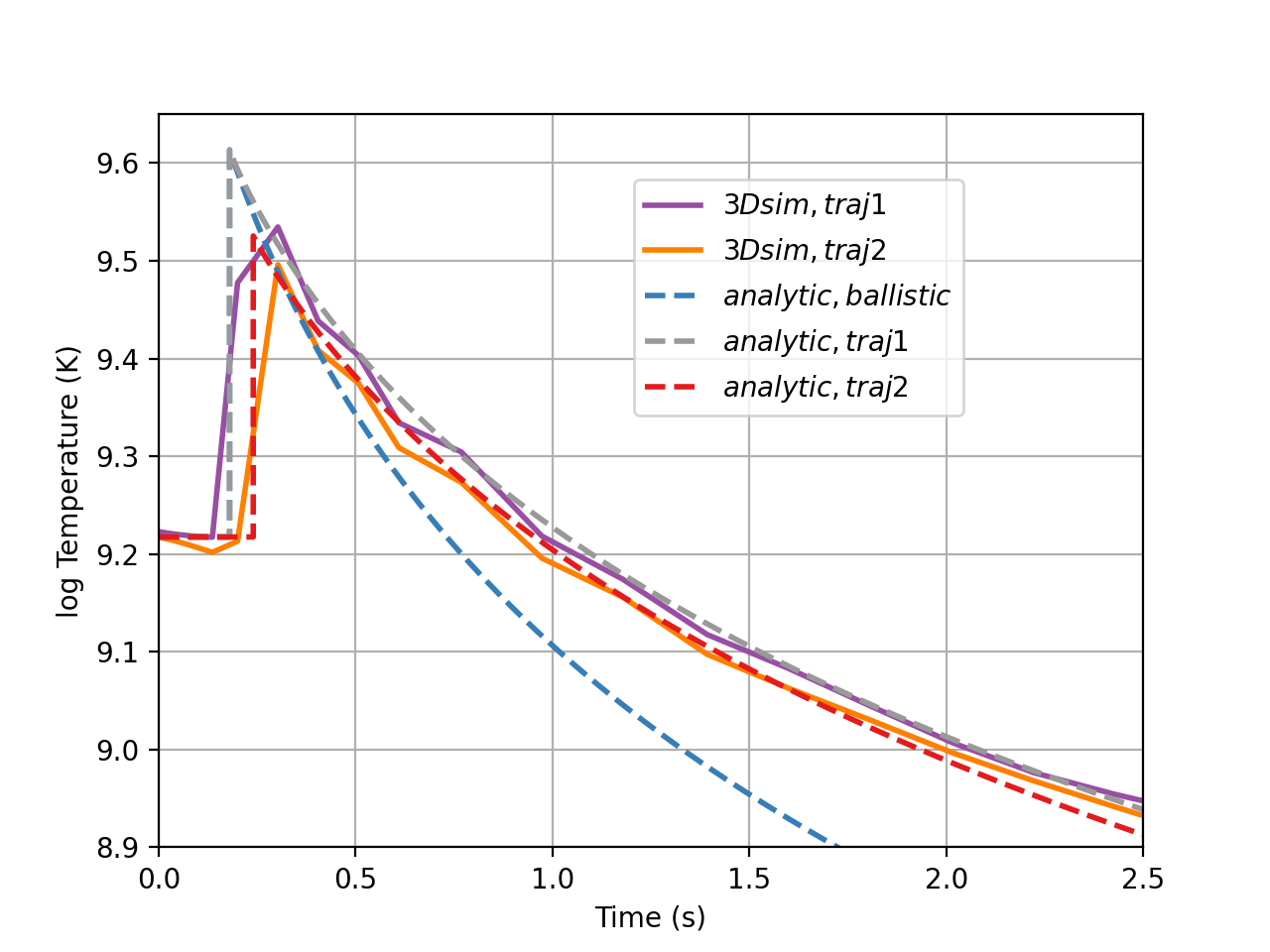}
    \caption{Temperature versus time for trajectories of carbon-shell material from 3-dimensional Smooth Particle Hydrodynamics simulations~\citep{2020ApJ...895...82V} of an asymmetric supernova (solid lines).  The dashed curves show analytic solutions of these same trajectories.  The simulated results underestimate the peak temperature for these models by, in some cases, more than 50\%.  However, the analytic model assuming ballistic trajectories, a common assumption, does not capture the deceleration, cooling the ejecta too quickly.  A deceleration term is required to match the simulated results.  Note that Smooth Particle Hydrodynamics and Eulerian codes typically perform equally poorly in capturing the peak shock temperature~\citep{2006ApJ...643..292F}}
    \label{fig:trajcomp}
\end{figure}

{\it Advantages and Disadvantages of In-Situ Burning:}  Combining all of the physics is impossible to do fully in a post-process calculation.  For this reason, some groups have begun to calculate the yields using large networks in the explosion calculation itself~\citep[e.g.,][]{2021ApJ...921..113S,navo+2023}.  This captures many of the trajectory issues:  long-lived engines, 3-dimensional asymmetries and their resultant shock effects.  These simulations are pushing the limits of our computational abilities, limiting the resolution of the simulations.  This low-resolution means that multi-dimensional models struggle to resolve the shock and the peak shock temperature in these calculations tend to be lower than predicted by resolved 1-dimensional or analytic solutions~\citep[for example, see the Sedov test models for the FLASH code][]{2002ApJS..143..201C}.   Figure~\ref{fig:trajcomp} demonstrates some of the effects discussed here by comparing analytic to simulated (3-dimensional) results.  Here we consider 2 different trajectories:  material in the carbon layer being shocked in in a 3-dimensional supernova explosion.  The simulated runs come from a Smooth Particle Hydrodynamics simulation of an asymmetric explosion mimicking an asymmetric but short-lived explosion~\citep{2020ApJ...895...82V}.  Comparing analytic and simulated trajectories in this figure demonstrate both the resolution limitations of 3-dimensional runs and the too-simple approximations of an analytic solution assuming ballistic trajectories.

Further, as we have demonstrated, the turbulent mixing of material is typically low.  3-dimensional simulations can not capture the turbulent mixing for the high Reynolds numbers ($Re>1000$) in supernovae where the number of grid points along a single dimension required to resolve the simulation is approximately $2 Re^{3/2}$~\citep{2003JTurb...4...22J}.  This would underestimate the mixing.  However, the numerical diffusion in these calculations far exceeds the actual mixing and the amount of mixing in these calculations is overestimated greatly. This could be addressed by using a lagrangian method (e.g. smooth particle hydrodynamics) or employing a steepening factor in the advection scheme \citep[e.g.,][]{plewa1999}

Finally, we should note that extracting Lagrangian tracer particles from 3D models is also a source of uncertainty when calculating the nucleosynthesis in a post-processing step \citep{harris+2017,sieverding+2023}. In particular, Lagrangian tracer particles tend to under-represent low-density regions, which can affect the predicted yields of alpha-rich freeze products, such as $^{44}$Ti, by roughly an order of magnitude \citep{harris+2017,navo+2023}.  

Ultimately, accurate yield estimates will require combining analytic, 1-dimensional and multi-dimensional calculations with in-situ and post-process nuclar burning.   

\subsection{Electron Fraction and Neutrino Effects}

As we shall see with our example study of our innermost ejecta, isotopic yields can be very sensitive to the electron fraction, $Y_e = n_e/(n_p+n_n)$, the ratio of the number of electrons to the number of protons plus neutrons of the ejecta.  Core-collapse engine modelers have been steadily improving the fidelity of their physics, e.g. behavior of dense nuclear matter, neutrino properties (including neutrino oscillations), to produce increasingly accurate solutions to the electron fraction of the ejecta~\citep{2017ApJ...839..132T,2020PhRvD.101d3009J,2020PhRvD.102j3017J,2021PhRvD.104h3025N,2023PhRvD.107j3034E,2023PhRvL.131b1001F,2025PhRvD.111l3038C,2025arXiv250916306R}.  Constraining the electron fraction of the ejecta from supernovae remains an active area of research and it remains a major uncertainty in the final ejecta composition.  As we shall see in our study of the innermost ejecta, the $^{44}$Ti yield is extremely sensitive to value of $Y_e$.

\begin{figure}[!ht]
  \centering
  \includegraphics[width=0.48\textwidth]{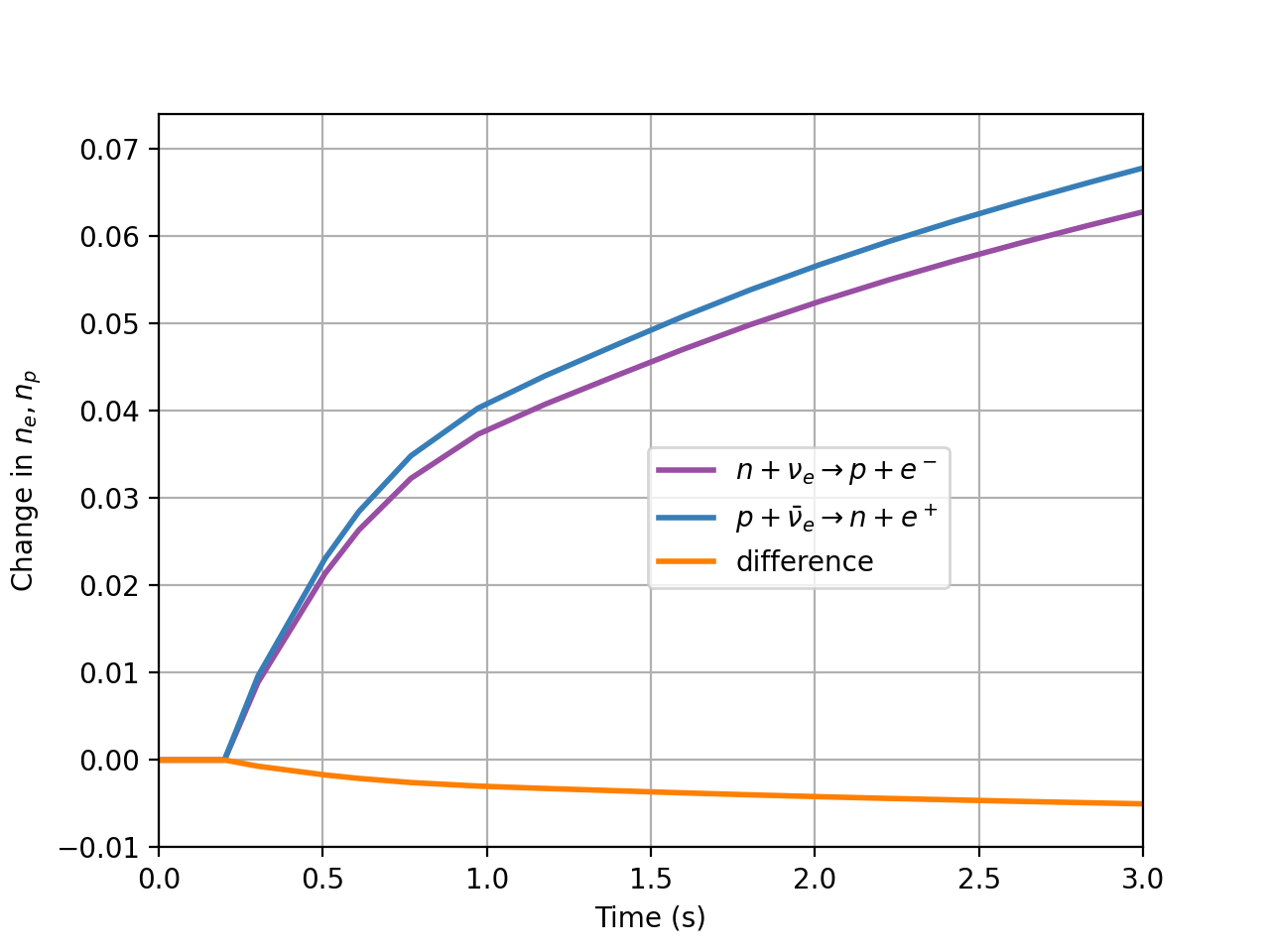}
    \caption{Change in neutron and proton fractions from electron neutrino and anti-neutrino absorption in a trajectory from material ejected near the supernova engine using neutrino luminosities from supernova core-collapse models.  Although the electron neutrino luminosities tend to be higher, their average energy is lower.  Hence, the net effect of neutrino absorption is to decrease (by less than 0.5\%) the neutron fraction (increasing the electron fraction).  Note that this model just focuses on the change of neutron and proton fractions and may change the over-all lepton fraction by much less than 0.5\%.}
    \label{fig:neut}
\end{figure}

Even if we perfectly calculate the electron fraction of the explosive engine, neutrinos can continue to evolve this fraction as the matter propagates through the star.  Oftentimes, to evolve the ejecta out to the edge of the star, simulations cut out the proto-neutron star to follow the ejecta to late times.  In many cases, the effect of neutrinos on the electron fraction on the ejecta after this ejecta moves beyond some radius is neglected.  We can estimate this error again using our trajectories from our carbon-shell shock heating model.  Figure~\ref{fig:neut} shows the change in the neutron and electron number fractions as a function of time (the net change is the difference) assuming the following neutrino properties: $L_{\nu_e} = 10^{57} \nu/s$ with average energy $E_{\nu_e}=10 {\rm \, MeV}$ and $L_{\bar{\nu}_e} = 7.5 \times 10^{56} \nu/s$ with average energy $E_{\bar{\nu}_e}=12 {\rm \, MeV}$.  Although the interaction rate is not negligible ($>5\%$), the net change in the $Y_e \sim 0.5\%$.  Given our assumption of a large free proton/neutron fraction, this is likely to be an overestimate on the effect of neutrinos.  This effect is much lower than the uncertainties in the neutrino effects during the engine and the supernova community has correctly focused on the neutrino effects in the engine.

\subsection{Tying to Observations}

The different supernova yields tie to a range of potential observations:  galactic chemical evolution, yields in supernova remnants, and spectra in the supernova itself.  In all cases, we have observations from atomic lines (which typically can't differentiate the isotopic abundances and rely upon knowing the excitation and ionization state of the atoms) and nuclear decay (which provide more direct probes).  Here we briefly discuss the uncertainties of these different observations.

{\it Galactic Chemical Evolution:}  Astronomers now have a growing number of observations studying the detailed yields in stars from which we can study the evolution of the elements in the Milky Way as a function of cosmic time~\citep{2020ApJ...900..179K}.  In addition, observations of decay photons from long-lived radioactive nuclei can provide insight into the nature of supernovae in the last few million years~\citep{2019BAAS...51c...2T,2021PASA...38...62D}. These studies can place constraints on generic trends for each of the sources (supernovae, neutron star mergers, novae, stellar winds) but our lack of understanding of the initial mass function of stars, individual supernova properties, mixing within the galaxy, etc. make it very difficult to probe conditions of the supernova or the underlying physics from these observations.  

{\it Supernova Remnants:}  In supernova remnants, we study a single supernova event, avoiding many of the population uncertainties in galactic chemical evolution.  The expansion of the ejecta can provide a rough handle on the strength of the explosion.  For a handful of remnants, we have detailed X-ray observations with line features that can be used to estimate the ejecta masses.  The difficulty with these X-ray lines is that a) X-ray lines typically do not differentiate between different isotopes, b) we only observe the X-rays if the material is shock heated from the deceleration of the shock as it propagates through the surrounding medium (between the forward and the reverse shock) and c) we do not know the excitation/ionization state of the shocked ejecta.  

Shock heating incurs a number of uncertainties because the nature of the heating depends sensitively on the conditions of the surrounding medium.  In massive stars, this is typically set by mass loss episodes in the final stages of the star's life prior to collapse.  It is known that the simple steady-state wind solution is far from correct:  mass loss is neither symmetric nor steady-state~\cite[for a review, see][]{2020ApJ...898..123F}.  Strong shock heating only occurs when considerable mass is swept up and it is often difficult to determine whether the observed element was produced in the explosion or was part of the star or swept up in the explosion.  Observations of the X-ray emission rely on models to both determine what mass has yet to be shock-heated and what fraction of the shock-heated material is swept up material versus material produced in the explosion.  For example, iron observed in supernova remnants includes both stellar iron (e.g. at formation, a 20\,M$_\odot$, solar-metallicity star contains $\sim 0.03$\,M$_\odot$ of iron) and iron in the interstellar medium.  For old remnants that have swept up considerable interstellar medium material, the bulk of the observed iron is likely to arise from swept up material.  The ``at-formation'' and interstellar medium iron must be removed from the total budget to determine the iron produced (via decay of $^{56}$Ni) in the explosion.

In addition, the remnant ejecta are almost certainly not in local thermodynamic equilibrium.  In this case, inferring the abundance fractions from the line strengths can be quite uncertain.  Typically, scientists studying remnants assume that different elements are in similar out-of-equilibrium states and then, perhaps, the ratio of different element abundances can be calculated~\citep{2023MNRAS.525.6257B}.  But the physics of the excitation levels of the atom is much more complex than this simple picture and is difficult to predict the uncertainties in the line features.

Gamma-ray observations of supernova remnants provide a much more direct probe of supernova engine.  Radioactive isotopes decay whether or not the material is shock heated and, especially at the low densities of these remnants, the $\gamma$-ray emission from decay can be directly tied to the amount of that isotope.  For example, by observing the emission lines from $^{44}$Ti decay, astronomers were able to decisively prove that, at least for some supernova explosions (i.e. the explosion behind the Cassiopeia A supernova remnant) are produced by a convective engine~\citep{2014Natur.506..339G,2017ApJ...834...19G}.  There are a few uncertainties and limitations tied to the observations:  e.g. the decay lines from $^{44}$Ti in the hard X-ray push the limits on what NuSTAR can do, the number of supernova remnants that can be studied in $\gamma$-rays is currently limited to 2:  Cassiopeia A and SN 1987A.  But the accuracy of these observations ensure they dominate the constraints on studies of the supernova engine.

\begin{figure}[!ht]
  \centering
  \includegraphics[width=0.5\textwidth]{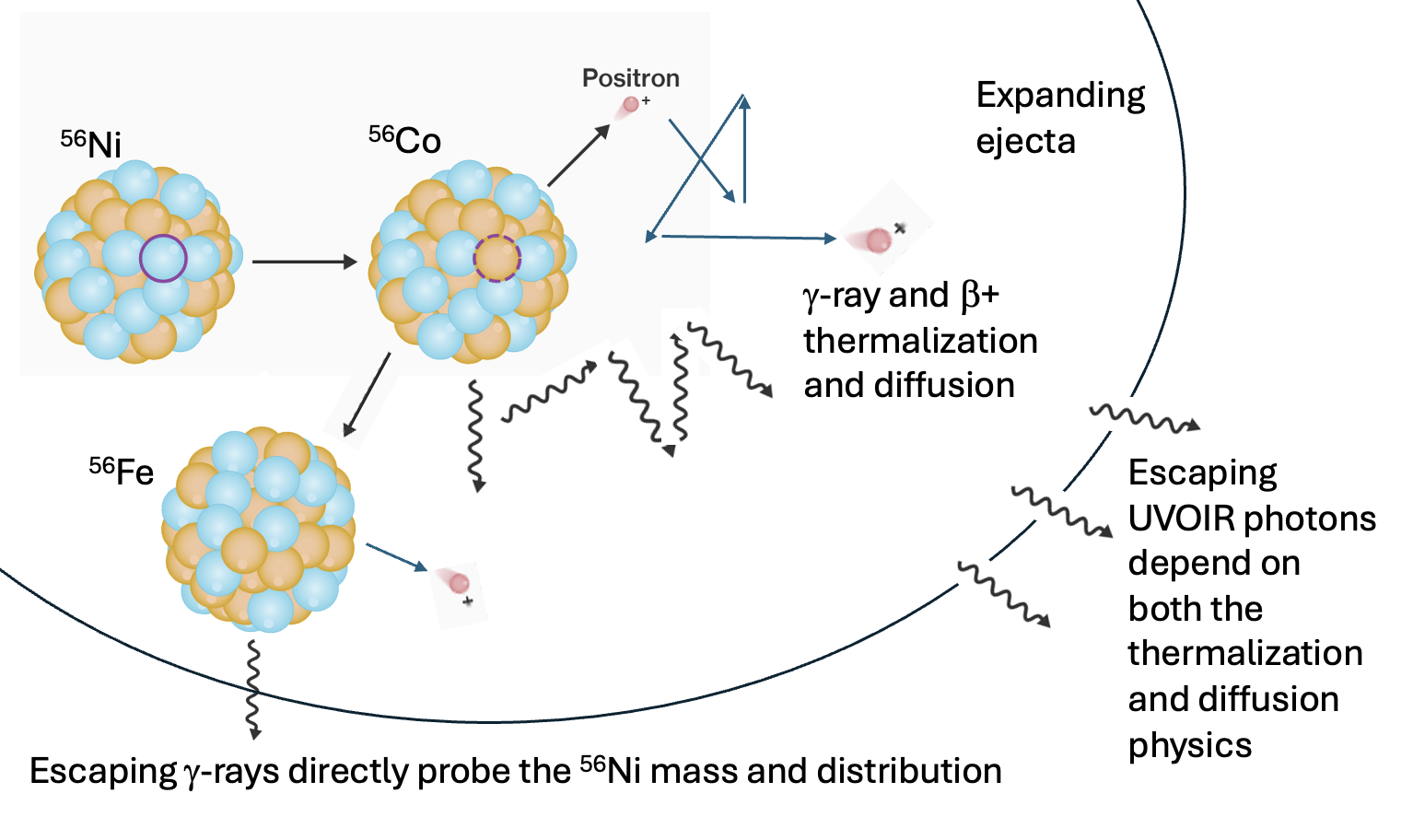}
    \caption{Diagram comparing the uncertainties in observations of supernovae.  Even if we know the power source of the emission, the infra-red through ultraviolet photons face a number of uncertainties. In the case of a radioactive decay energy source, the $\gamma$-ray photons and charged particle energy released in the decay must be re-processed to be observed.  The observations depends sensitively on the reprocessing and the atomic physics driving this.  For $\gamma$-rays, we observe the decay photons directly.}
    \label{fig:diagram_nuc}
\end{figure}

{\it Supernova Transients:}  Observations in the spectra of the transient itself has the potential to be the most powerful probe of nucleosynthesis in supernovae.  Like supernova remnants, we have a broad set of multi-messenger signals of these transients.  For example, for a growing list of supernovae, we have previous observations of the supernova progenitor, placing constraints on the initial conditions~\citep{2009ARA&A..47...63S}.  Combined with spectra from the supernova ejecta during the explosion, these observations can provide incredible insight into the progenitor star and the explosive mechanism (e.g. ejecta masses, velocities and yields).  However, the bulk of the observations lie in the infra-red and optical bands and it is difficult to tie these observations directly to the physical properties of the explosion.  One of the major uncertainties is the fact that we don't fully understand the dominant power source for the emission in infra-red and optical bands.  Internal shocks, external shocks, decay of radioactive isotopes and long-lived central engines can all contribute to the observed emission.  Different energy sources can produce very similar light-curve properties~\citep{2025arXiv250115702N}.  These degeneracies arise because the infra-red through ultraviolet emission arises primarily from processed light (Figure~\ref{fig:diagram_nuc}).  Even for simple explosion models, different codes produce very different results in many bands~\citep{2022A&A...668A.163B}.   

Spectra in the infra-red through ultraviolet range are more constraining than light-curves alone, but face their own challenges.  At early times, the spectra are dominated by material above the photosphere (providing only a partial window of the yields).  At late times, a more complete analysis can be done but, just as with supernova remnants, out-of-equilibrium effects limit what we can infer from the observations.  Detailed, out-of-equilibrium models are required to infer detailed yields from these observations.  In addition, constraining the supernova engine through atomic spectra typically is done through late-time observations of iron.  Here again, any analysis must disentangle the $^{56}$Ni produced in the explosion from the iron in the star.

On the other hand, $\gamma$-rays from radioactive decay provide a more direct probe of a subset of the yields:  radioactive isotopes with half-life timescales on par with the explosion~\citep{2020ApJ...890...35A}.  The $\gamma$-rays from SN 1987A both led to the initial argument for mixing in the supernova engine~\citep{1988ApJ...329..820P} and ultimately placed strong constraints on the distribution of this $^{56}$Ni~\citep{2003ApJ...594..390H,2005ApJ...635..487H}.  High energy $\gamma$-rays have similar interactions with bound or free electrons, reducing the uncertainties due to out-of-equilibrium features of the atoms.  In addition, the observed photons only undergo a few scatterings (Figure~\ref{fig:diagram_nuc}) and the resultant transport uncertainties are minor (compare the differences in \cite{2004ApJ...613.1101M} to the light-curve comparisons in ~\cite{2022A&A...668A.163B}).  With upcoming pathfinder missions like COSI~\citep{2024icrc.confE.745T}, $\gamma$-ray observations will be able to probe the yields at larger distances and higher precision than has been done before, increasing the potential of this probe of the supernova engine.

\section{Nuclear Physics Uncertainties}
\label{sec:nuclear}
For given astrophysical conditions, the nucleosynthetic yields of core collapse supernovae depend on a complex network of nuclear reactions \citep{The1998,Magkotsios2010}.  
The primary interest for the production of observable $\gamma$-ray emitters such as $^{44}$Ti is nucleosynthesis in the innermost ejecta that depends on nuclear reactions involving nuclei ranging from the triple-alpha reaction to isotopes in the Si group and Fe groups \citep{The1998,Magkotsios2010,Hermansen2020,Subedi2020}. As $Y_{\rm e}$ is close to 0.5, reactions proceed around isotopes with equal neutron and proton numbers ($N \approx Z$) and thus, for $Z>20$, along the neutron deficient side of the valley of stability involving both stable and unstable nuclei. Some reactions of lighter nuclei are also critical, such as 3$\alpha$, $^{12}$C($\alpha,\gamma$)$^{16}$O.
%, however, their rates tend to have relative lower uncertainties and thus tend to contribute less to the nuclear uncertainty budget. 

For typical (but not all) thermodynamic trajectories, material is rapidly heated to temperatures in the 5-10~GK range \citep{Magkotsios2010}. At such temperatures, initially-present nuclei synthesized during the preceding stellar evolution are mostly destroyed by photo-disintegration. At this stage, the composition is determined by the equilibrium between rapid forward and reverse reactions, and approaches nuclear statistical equilibrium (NSE). In equilibrium, the composition is independent of individual reaction rates and depends solely on nuclear masses and partition functions. Depending on the conditions, the equilibrium composition tends to be a mixture of lighter particles (protons, neutrons, and $\alpha$-particles) and heavier nuclei. As the material cools, the strongly temperature-dependent charged particle reaction rates decrease and more and more reactions drop out of equilibrium. This results in the breakdown of the equilibrium into smaller interconnected clusters of equilibrium:  QSE - Quasi-Statistical Equilibrium~\citep{Meyer1998}. As time goes on, more and more clusters break up into smaller clusters until reactions connect individual nuclei and then freeze out. Key nuclear reactions tend to be reactions that connect equilibrium clusters (or individual nuclei out of equilibrium during later times of the process) at critical phases of the process, thus governing the buildup of heavier elements. 

The important reaction rates from Si to Ni are generally not as well studied experimentally as reactions in lighter mass regions. This creates significant nuclear reaction rate uncertainties. Even for stable nuclei, there are gaps in experimental data that can affect the predictive power of models. At the lower part of the relevant temperature range of a few GK, cross sections can be in the micro-barn range or even lower. Direct measurements of reaction cross sections are therefore challenging even for reactions on stable nuclei where accelerators can produce relatively intense beams. For unstable nuclei, available beam intensities are typically much lower. Some direct measurements have been performed on $^{44}$Ti, which has a sufficiently long (59 years) half-life for samples to be produced that can then serve as targets or beam sources \citep{Sonzogni2000,Margerin2014,Ejnisman1998}. No direct measurements have been performed so far for many important reactions on unstable nuclei for CCSN nucleosynthesis in the Si-Ni range because of the lack of sufficiently intense radioactive beams. These reactions are marked in Fig.~\ref{fig:HFAllVariations}. Such measurements are a future goal for new radioactive beam facilities. 

The majority of reaction rates in current nuclear networks therefore need to be predicted from theoretical models. However, these models do not reach the required accuracies. The shell model can, in principle, predict individual resonances for proton capture reactions. Such calculations are available, for example, for the potentially important $^{47}$V(p,$\gamma$)$^{48}$Cr, $^{45}$Cr(p,$\gamma$)$^{46}$Mn, and $^{46}$Cr(p,$\gamma$)$^{47}$Mn reactions \citep{Fisker2001}. However, in this intermediate mass range uncertainties are largely due to the limited model space, in particular for predictions of excitation energies and can result in orders of magnitude uncertainties in the predicted reaction rates (see for example \citet{Langer2014}). 

For the majority of reaction rates the statistical Hauser-Feshbach model is employed \citep{Rauscher2000,kawano+2019,koning+2023}. Statistical models suffer from uncertainties of optical model potentials and statistical nuclear structure properties such as level densities and $\gamma$-ray strength functions. In some cases, statistical approaches may break down all-together due to low level densities, especially closer to the proton drip line. Masses and partition functions are also needed to determine equilibrium abundances. While the relevant masses have been determined experimentally with sufficient precision, the partition functions depend on the spins and excitation energies of low-lying states where there may be remaining uncertainties for unstable nuclei. For example, the low lying spins of $^{45}$V remain uncertain. The properties of these low-lying states are also needed to determine stellar enhancement factors, i.e. corrections to experimentally determined ground-state reaction rates due to thermal population of low-lying (below a few hundred keV) excited states of interacting nuclei in the stellar environment. 

In addition, while the initial pre-explosion composition is largely erased by the initial heating for many thermodynamic trajectories, this is not true for all conditions. Furthermore, the thermodynamic conditions themselves are sensitive to the pre-supernova structure of the star. For these reasons, nuclear reactions during stellar evolution that impact the pre-supernova structure add additional nuclear uncertainties that need to be addressed.

In the following, we discuss theoretical (Section.~\ref{sec:HFVariations}) and experimental (Section~\ref{sec:ExpConstraints}) nuclear uncertainties in more detail. Additional uncertainties from nuclear network implementations are discussed in Section \ref{sec:nuclear_networks}. Of key importance is the determination of the critical nuclear reaction rates, for which we discuss methods in Section~\ref{sec:SensitivityStudies}. For these nuclear reaction rates, uncertainties need to be carefully estimated, which can then be propagated to observable astrophysical yields (Section~\ref{sec:NucErrorPropagation}). This provides the nuclear contribution to the total error budget of supernova nucleosynthesis models and can be used to identify, for specific observables and models, the dominant nuclear uncertainty contributions that need to be addressed by experiments or improved theory.

\subsection{Uncertainties in Theoretical Reaction Rates}
\label{sec:HFVariations}
\begin{figure}
  \centering
  \includegraphics[trim={0cm 0.6cm 1cm 1cm},clip,width=0.5\textwidth]{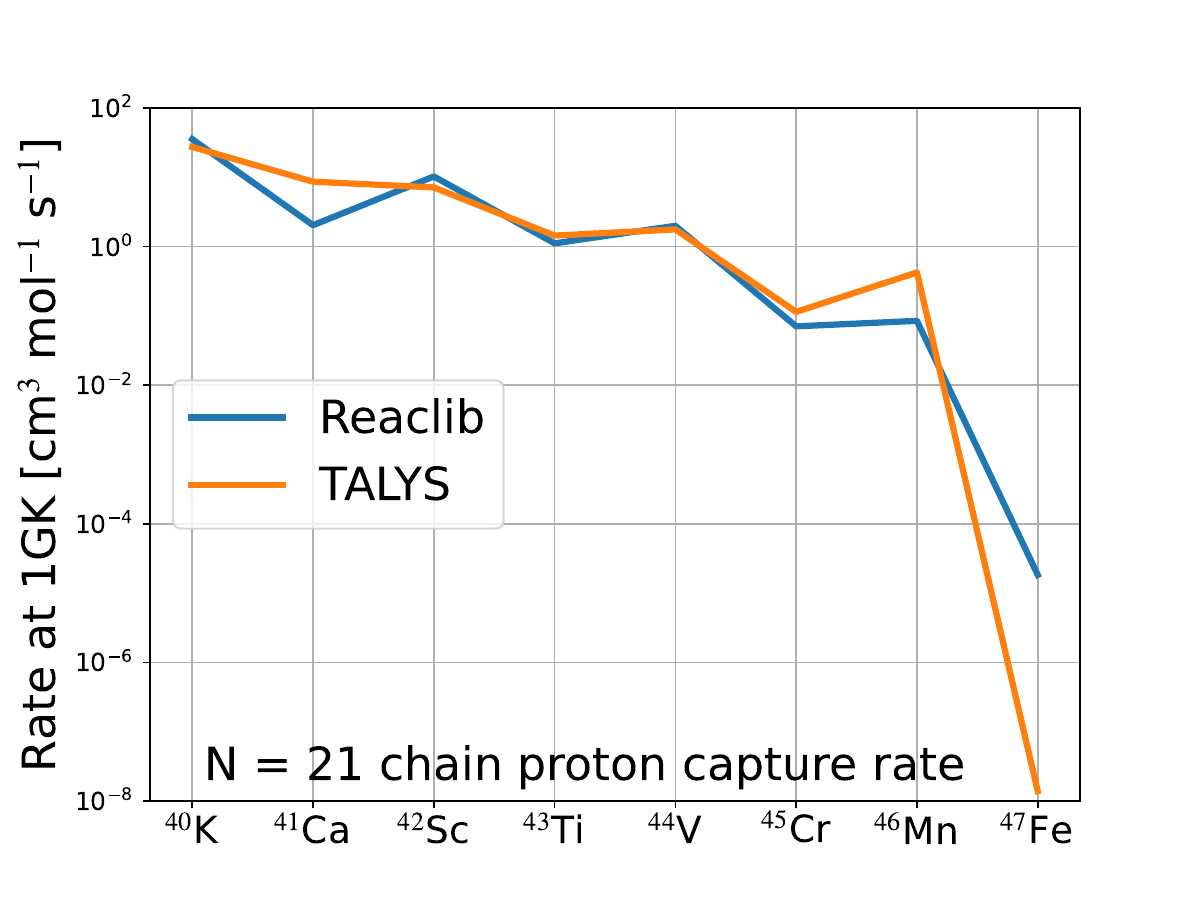}
  \caption{Proton capture rates ($(p,\gamma)$) for the $N=21$ isotonic chain from Reaclib library (blue) and TALYS with default options (orange) at 1 GK.}
  \label{fig:HF_model_variation}
\end{figure}

The statistical Hauser-Feshbach model is used to predict the majority of experimentally unconstrained nuclear reaction rates. Fig. \ref{fig:HF_model_variation} shows a comparison of $N=21$ isotopic chain proton capture reaction rate predictions from commonly employed Hauser-Feshbach codes: NON-SMOKER \citep{Rauscher2000} as included in the Reaclib library \citep{cyburt2010jina} and TALYS \citep{koning+2023} with the default settings in the latter. The models follow closely throughout proton richness scale and show similar odd-even features. At low neutron numbers, i.e. closer to stability, the differences in the capture rates are smaller than a $\mathcal{O} (0.5)$, however as we move away from available experimental data to $^{47}$Fe the models differ by as much as $\mathcal{O} (1.5)$.
%\textcolor{red}{ ADAPT TO NEW FIGURE The three models follow each other closely throughout neutron richness scale and show similar odd-even features. At low neutron numbers, i.e. closer to stability, the differences in the capture rates are smaller than a $\mathcal{O} (1)$, however, with increasing neutron numbers, as we move away from available experimental data, the models start differentiating by as much as a few orders of magnitude.}

Quantifying the uncertainties for these rate predictions poses various challenges. There are two principal components to the uncertainty: (1) the intrinsic statistical model accuracy limited by the basic model assumptions inherent in the statistical average treatment of nuclear properties, and (2) uncertainties in the predictions of the statistical nuclear properties and optical model potentials needed in the statistical model, primarily nucleon and $\alpha$ optical model potentials, nuclear excitation level densities, and $\gamma$-ray strength functions for the target nucleus and final nuclei in all energetically possible exit channels. While (1) is likely to a large extent to be a statistical and thus uncorrelated error, uncertainties in (2) are highly correlated affecting all nuclei in a systematic way.

Intrinsic uncertainties are difficult to estimate given the scarcity of experimental data. Some guidance can be provided from neutron capture rate predictions, as a large data set of relatively accurate experimental nuclear data at astrophysical energies is available. Optical model potentials, level densities, and gamma-strength functions are constrained at least to some extent by experimental data as well. Comparisons of predictions of neutron capture rates on stable nuclei with experimental data indicate typical uncertainties of factors of 2 \citep{Beard2014}. Away from stability this intrinsic uncertainty may further increase. As one approaches the proton drip line, level densities decrease limiting the applicability of the statistical model, especially for the lowest temperatures of interest \citep{Rauscher:1997,Subedi2020}. In addition, for $\alpha$-induced reactions $\alpha$-clustering effects may also impact reaction rates \citep{Wiescher2025}.

\begin{figure}
  \centering
  \includegraphics[width=0.5\textwidth]{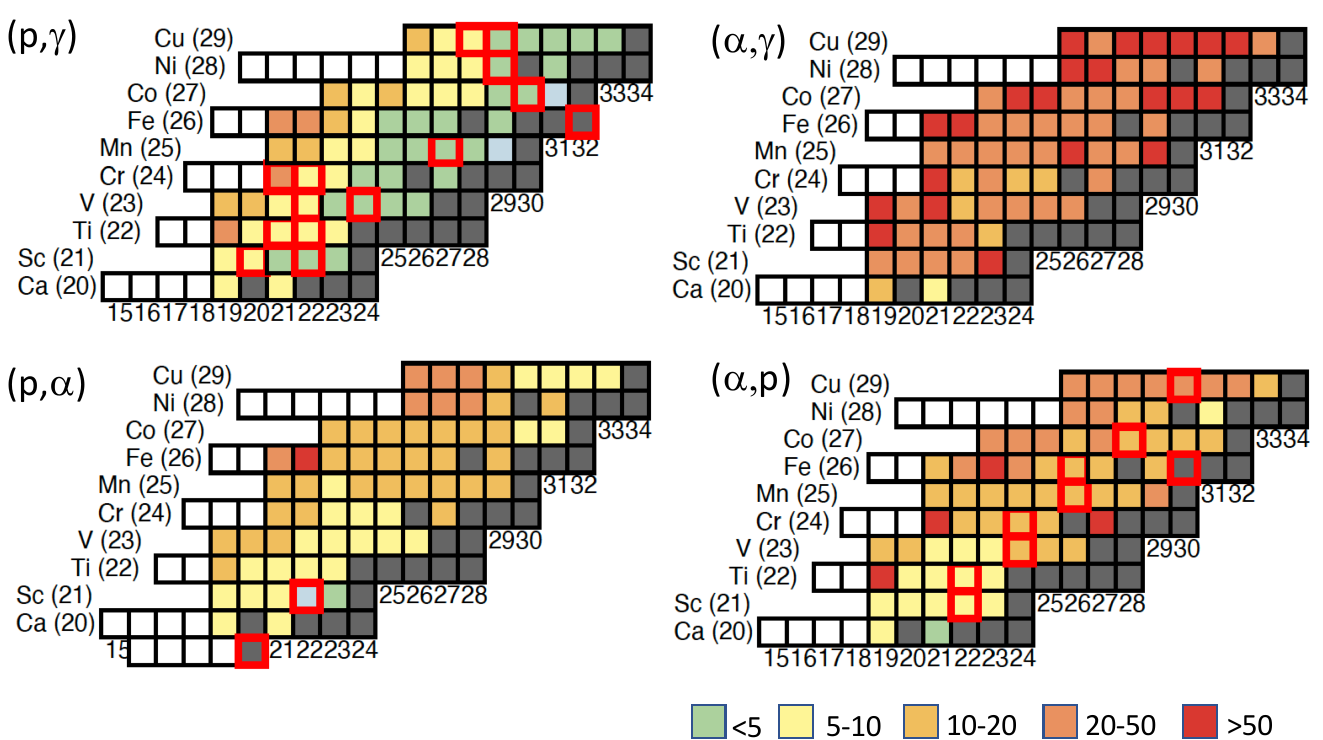}
  \caption{Maximum variations in the theoretical Hauser-Feshbach reaction rates that arise from assumptions in the physics model choices available
    in the \texttt{TALYS} code. Potential key reactions identified in previous work are marked as red squares. }
  \label{fig:HFAllVariations}
\end{figure}

Uncertainties from statistical model ingredients are also difficult to assess. Recently a theory based uncertainty quantification has been performed for the Koning-Delaroch optical model potential \citep{2023PhRvC.107a4602P}.  However, strong theory constraints on the uncertainties of level densities and gamma-strength functions are still missing. One commonly employed approach is to implement a variety of available models of optical model potential, level density, and gamma-strength functions. The TALYS code is especially well suited for such studies as it readily provides multiple model options for each of these components. This approach has strong limitations. On one hand, this may only provide a minimal range of possible predictions, as the range is defined by the selection of models that happen to exist and be made available by the authors of the code. In addition, uncertainties from the models themselves are neglected. True uncertainties may therefore be larger than expected. On the other hand, the model selection may include models that are less favored by experimental data or contain questionable theoretical assumptions for a specific mass region, thus artificially inflating uncertainties. Nevertheless, this approach can provide a first estimate of the possible range and sensitivities and has commonly been employed in other areas of nuclear astrophysics \citep{Liddick2016, Pereira2016}. Fig.~\ref{fig:HFAllVariations} shows resulting reaction rate uncertainties $\delta r$, calculated as 
%\begin{equation}
$\delta r = \max( r_{i {\rm; high}}/r_{i {\rm; low}})$
%\end{equation}
for all temperature points $i$ within 1-5~GK. While uncertainties are relatively small near 
stability, uncertainties dramatically increase away from stability, similar to what is found for neutron capture rates \citep{Liddick2016}. While for proton capture rates, uncertainties reach factors of 20-50, uncertainties of more than a factor of 50 are common for $\alpha$-capture reactions. This is due to the well-known uncertainties in $\alpha$-nucleus optical model potentials \citep{Mohr2015}. There has been some convergence on $\alpha$-nucleus optical model potentials in the $A=20-50$ mass region in terms of achieving agreement between model predictions and experimental data on stable nuclei \citep{Mohr2015}. Uncertainties near stability may therefore be somewhat overestimated in Fig.~\ref{fig:HFAllVariations}. However, it's unclear how this evolves further away from stability.

\begin{figure}
  \centering
  \includegraphics[width=0.5\textwidth]{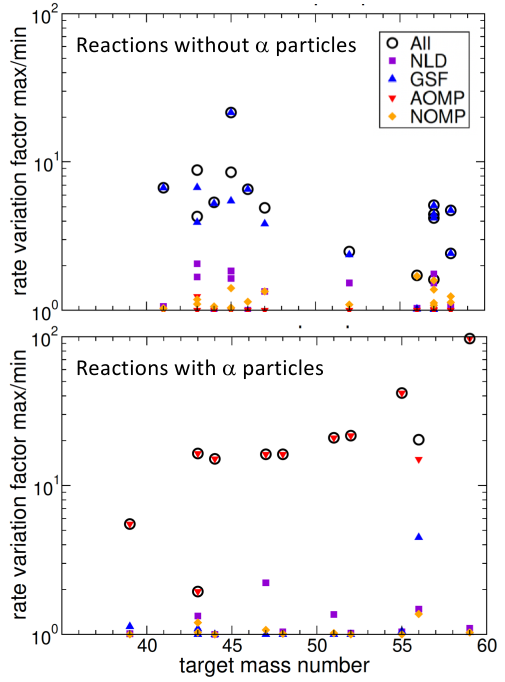}
  \caption{Maximum variations in the theoretical Hauser-Feshbach reaction rates 
    for temperatures in the 1-5~GK range arising from the physics model choices available
    in the \texttt{TALYS} code. The variations are for potentially important reactions that do not include (upper panel) and do include (lower panel) $\alpha$-particles displayed as function of mass number of the target nucleus. Shown are total variations (circles) as well as variations due to just level densities (violet squares), $\gamma$-strength (blue triangle up), $\alpha$-optical potential (red triangle down) and nucleon optical potential (orange diamond).}
  \label{fig:HFVariations}
\end{figure}

Fig.~\ref{fig:HFVariations} shows the same uncertainties for a subset of potentially important reactions, separated by the type of nuclear structure ingredient. The reactions are marked in Fig.~\ref{fig:HFAllVariations}, and include in addition $^{56}$Co(n,p)$^{56}$Fe and $^{57}$Ni(n,p)$^{57}$Fe. The $\alpha$-nucleus optical model potential variation dominates the uncertainty for $\alpha$-induced reactions, with other ingredients contributing mostly factors of 2 or less. For reactions that do not involve $\alpha$-particles, uncertainties range from factors of 1.6-20, and tend to be dominated by the $\gamma$-strength function. Level densities and optical potential parameters contribute at most a factor of 2 uncertainty.

\subsection{Uncertainties in Experimental Reaction Rates}
\label{sec:ExpConstraints}

Experimental constraints are critical to reducing nuclear uncertainties and, equally important, to reliably characterize theoretical uncertainties. For most cases, experimental data and theoretical information need to be combined to obtain a reliable astrophysical reaction rate. Even if experimental cross sections or resonance strengths are available, they rarely extend over the entire relevant temperature range. Theoretical calculations can then be used to extrapolate data. In addition, experimental information is usually limited to reactions on the target in the ground state. Theory is needed to predict corrections for reactions on thermally excited states, the so-called stellar enhancement factor. For reactions with no experimental information, experimental constraints on reactions nearby can still provide important guidance to quantify uncertainties in rates determined from theory. Given the scarcity of relevant nuclear data, more experimental data are therefore needed to broadly reduce nuclear uncertainties in core collapse supernova nucleosynthesis calculations. Nevertheless, important work has been done on a few critical reactions. Here, we identify two illustrative reactions for which the rate can be experimentally constrained: $^{44}$Ti($\alpha$,p)$^{47}$V and $^{23}$Na($\alpha$,p)$^{26}$Mg (important for other $\gamma$-ray emitters identified by \cite{Hermansen2020}). These cases are taken as case studies.

%\begin{description}
%\item[Hauser-Feshbach rates on stable nuclei] Collect experimental
%  data. If there is enough data to determine the reactions rate, use
%  the experimental rate directly. If not, the theoretical
%  Hauser-Feshbach reaction rate can be tuned to match the data, then
%  subsequently used to determine the rate over a broader range of
%  temperatures. Care must be taken to extrapolate the model-based
%  uncertainties in this case.
%\item[Hauser-Feshbach charged particle rates on unstable nuclei]
%  Survey known cross sections for all reactions in the relevant energy
%  and mass range for explosive Si burning with Hauser-Feshbach
%  predictions to better understand its reliability for stable nuclei.
%  Described in more detail in Sec.~\ref{sec:HFVariations}.
%\end{description}

\paragraph{The $^{23}$Na($\alpha$,p)$^{26}$Mg Reaction}

\begin{figure}
  \centering
  \includegraphics[width=0.5\textwidth]{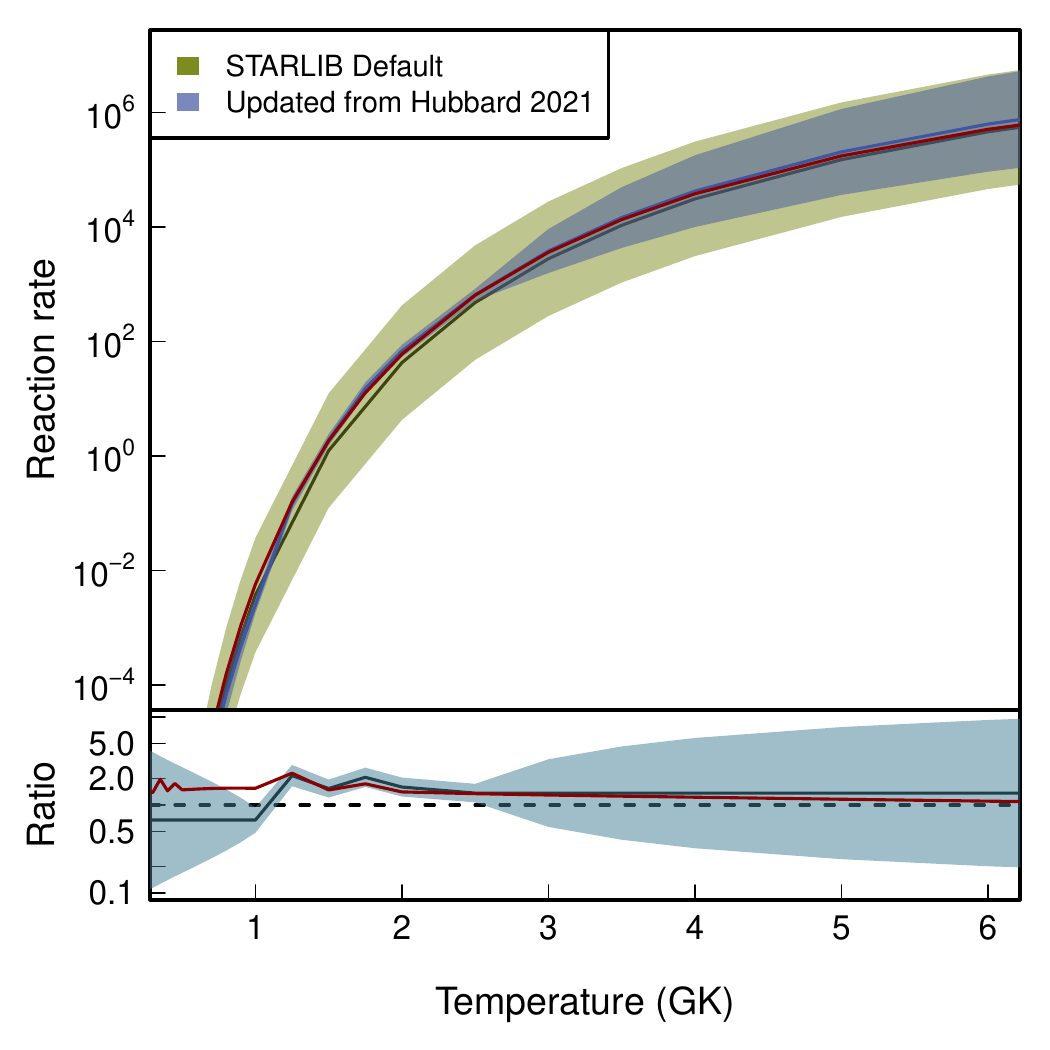}
  \caption{The rate for the $^{23}$Na($\alpha$,p)$^{26}$Mg reaction.
    Shown in green is the default \texttt{Starlib} reaction rate with its
    default uncertainty of a factor of 10, while the red curve shows the default \texttt{reaclib} reaction rate. In blue is the
    experimentally constrained reaction rate from \cite{Hubbard2021}.
    Above and below the measured temperature region, the
    \texttt{Starlib} reaction rates are scaled to to experimental
    range. A factor of 10 uncertainty is assumed at 0.1 GK and 10 GK
    for the extrapolations, with the uncertainty scaling linearly with
    temperature.}
  \label{fig:23Na-ap-rate}
\end{figure}

The $^{23}$Na($\alpha$,p)$^{26}$Mg reaction cross section has been
measured several times with different experimental methods, by
\cite{Tomlinson2015}, \cite{Howard2015}, and \cite{Almaraz2015}. In
the measurement of \cite{Howard2015}, a larger acceptance
measurement was performed allowing them to measure angular
distributions.  In turn, these were subsequently used to correct the other two measurements whose results assumed an isotropic distribution of
protons. These results were evaluated as a whole by
\cite{Hubbard2021}, who presented an experimentally-constrained
reaction rate between 1 GK and 2.7 GK. The recommended reaction rate
is up to a factor of two higher than the \texttt{STARLIB} default
rate, but also contains considerably tighter uncertainties on the
order of 30\% as shown in Fig.~\ref{fig:23Na-ap-rate}. The default reaction rate from the \texttt{reaclib} library is also shown. The agreement between both theory curves and the experimental data in this case is striking, particularly given the low mass of the system. Nevertheless, the experimental constraints on the uncertainties are considerably improved using experimental data.

The challenge then becomes one of extrapolating the reaction rate
outside of the measurement range. In this case study, we simply scale
the Hauser-Feshbach rate to the experimental rate at each end point.
It is physically unrealistic that the uncertainty would immediately
become a factor of 10 at the next temperature grid point. It is
equally unrealistic that the extrapolation uncertainty would be as
small as the last experimentally constrained one. Here, we make the
conservative assumption that the rate uncertainty is a factor of 10
at the far ends of the reaction rate (i.e. 0.01 and 10 GK), and
perform a log-linear interpolation of the uncertainties, which is apparent in Fig. \ref{fig:23Na-ap-rate}.

\paragraph{The $^{44}$Ti($\alpha$,p)$^{47}$V Reaction}

The $^{44}$Ti($\alpha$,p)$^{47}$V reaction is known to be a key, uncertain reaction in the production of $^{44}$Ti during supernova explosions \citep[see, for example,][]{The1998}. To accurately determine the reaction rate at temperatures for CCSN (temperatures between about 2.5 to 5 GK), the cross section should be experimentally constrained over center-of-mass energies between about 2 and 7 MeV \citep{Hoffman2010}. Two experiments have been performed to address this energy range by \cite{Sonzogni2000} and \cite{Margerin2014}. The former experiment was conducted in inverse kinematics covering an energy range of 5.7 - 9 MeV using the Fragment Mass Analyzer (FMA) at the ATLAS facility to measure the production rate of the $^{47}$V reaction product. The second experiment performed at REX-ISOLDE also used inverse kinematics, but rather measured the resulting proton exiting a helium gas cell using silicon detectors. This latter experiment measured a single upper limit cross section at an energy of about 4 MeV in the center of mass. These experiments are both summarized in some detail by  \cite{Chipps2020}, who noted that the experiments are in poor agreement even after corrections are applied for experimental effects \citep{Murphy2017}. 

\begin{figure}
  \centering
  \includegraphics[width=0.48\textwidth]{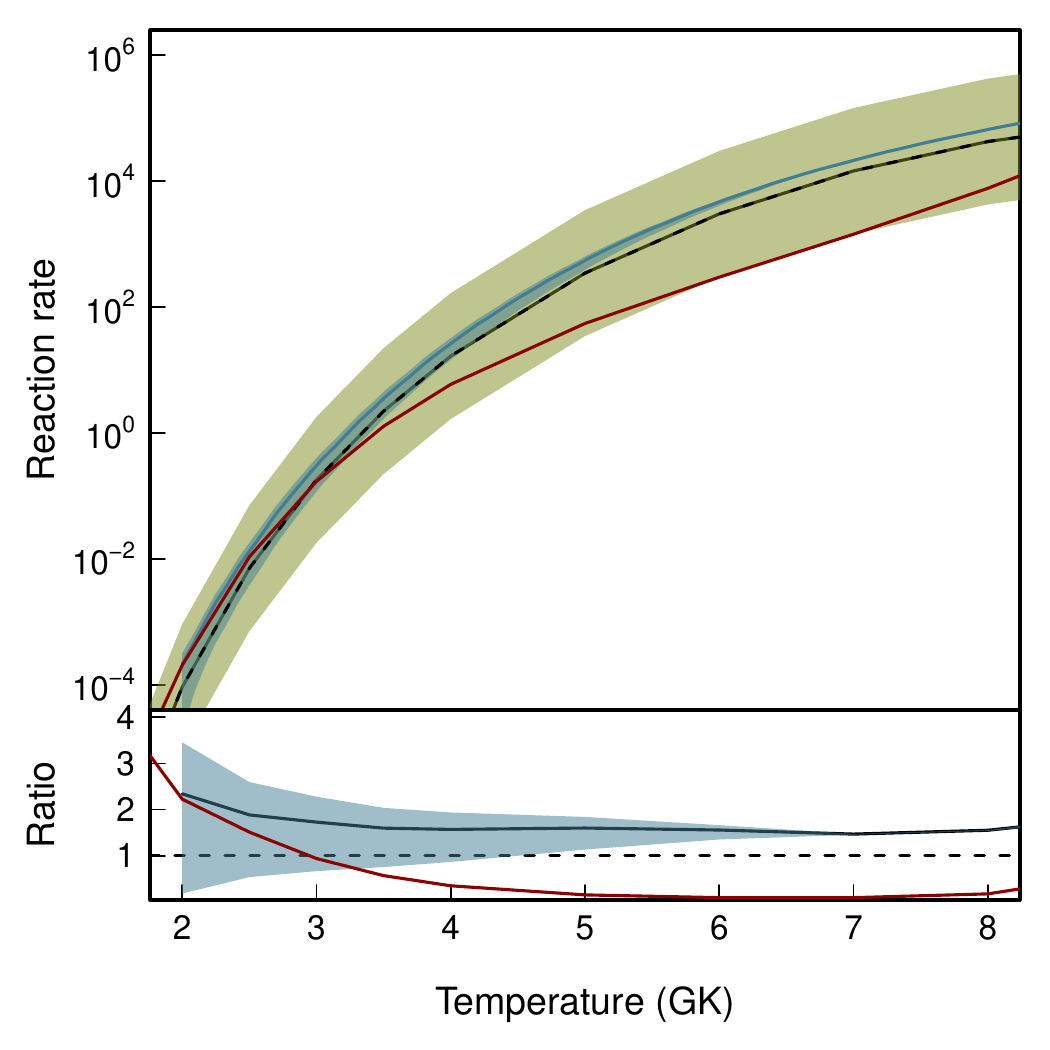}
  \caption{The rate for the $^{44}$Ti($\alpha$,p)$^{47}$V reaction.
    Shown in green is the default theoretical starlib reaction rate with its
    default uncertainty of a factor of 10, while in red the theory rate of \cite{cyburt2010jina} is shown. In blue is the
    experimentally constrained reaction rate from \cite{Chipps2020}.
    Above and below the reported temperature region, the
    \texttt{Starlib} reaction rates are scaled to experimental
    range. A factor of 10 uncertainty is assumed at 0.1 GK for the
    extrapolation, with the uncertainty scaling linearly with
    temperature.}
  \label{fig:Compare-TALYS-Chipps}
\end{figure}

A detailed discussion of the experimental data is provided by \cite{Chipps2020}, including warnings over the interpretation of extracted center-of-mass energies from the experimental data. Nevertheless, they used the experimental data to constrain theoretical Hauser-Feshbach calculations using the various $\alpha$-particle optical model potentials available in TALYS. The authors caution that their results depend on the validity of the Hauser-Feshbach statistical model over the entire energy range of interest, but provide a tabulated reaction rate and associated uncertainties between 2 GK and 10 GK. Although the available experimental data only directly constrain the cross section at temperatures above about 6~GK, they investigated the effect of extrapolating these experimentally-constrained rates using different model parameters in the TALYS code~\citep{koning+2023} and used their observed variations to recommend a reaction rate with uncertainties down to 2 GK. Temperatures below this value are not relevant for the conditions explored in the present study, but here we further extrapolate the rate below 2 GK to guarantee no numerical artifacts occur as the material cools. To achieve this, we scale the \texttt{STARLIB}  reaction rates to the lowest rate presented by \cite{Chipps2020}. To extrapolate the uncertainties, we scale the reported uncertainty at 2 GK log-linearly with temperature to a factor of 10 uncertainty at 0.1 GK. The resulting reaction rate in comparison to the default Hauser-Feshbach \texttt{Starlib} and reaclib rates is shown in Fig. \ref{fig:Compare-TALYS-Chipps}. The figure highlights three points: (i) that the default \texttt{Starlib} rate is less than a factor of two smaller than the experimentally-constrained one, ii) that the constrained rate uncertainty is significantly smaller than the factor of 10 assumed in \texttt{STARLIB}, and (iii) that there are significant differences between the Hauser-Feshbach reaction rates in \texttt{Starlib} and \texttt{reaclib}~\citep{cyburt2010jina}.

It should be stressed here that the reaction rate presented in \cite{Chipps2020} only includes statistical uncertainties from their normalization procedure and we use it here for illustrative purposes only. Future work should focus on folding model uncertainties with the experimental data by fully accounting for systematic experimental uncertainties, while also investigating the effect of different assumptions in Hauser-Feshbach codes. Examples of this using fully Bayesian analyses to take these effects into account for other astrophysical environments are presented in \cite{Moscoso2021, Odell2022, Iliadis2022}.

\paragraph{Conclusion and Outlook}

The experimental information constraining reactions impacting $\gamma$-ray emitters in supernovae is limited. Direct cross section measurements have only been performed for a few reactions. Often, as is the case for $^{44}$Ti($\alpha$,p)$^{47}$V, the experimental information is conflicting, so multiple measurements are required to reliably constrain the reaction rates. Furthermore, experimental cross sections are, so far, often limited to high energies where the cross section is large. These measurements are far from the energy ranges important during the supernova explosion, so theoretical models are needed to extrapolate to low energies. The accuracy of those extrapolations is, as yet, not well known.

With the recent development of more reliable, high-intensity radioactive ion beams and new detector technologies, the future for making more measurements of these cross sections directly at energies closer to those relevant for nucleosynthesis is bright. Below we discuss techniques that can help focus those experimental efforts to the most impactful reactions in the creation of $\gamma$-ray emitters from supernovae. On top of this, more systematic studies should be performed to understand the reliability of theory extrapolations from experimental data to astrophysical energies. This would be of great benefit in focusing experimental efforts in the most efficient manner.

\subsection{Nuclear Networks}
\label{sec:nuclear_networks}

\begin{figure*}
    \centering
    \includegraphics[width=\linewidth]{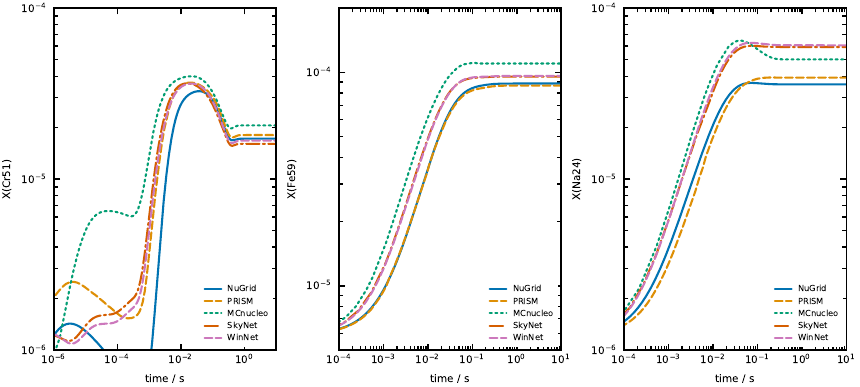}
    \caption{Shock-driven nucleosynthesis of radioactive isotopes in shock-expansion trajectories with carbon shell initial composition. From left to right, the trajectories represent $^{51}$Cr, $^{59}$Fe and $^{24}$Na production.}
    \label{fig:shell-network-comparison}
\end{figure*}

The evolution of the nucleosynthetic abundances are calculated by solving a series of ordinary differential equations that include the various nuclear reaction rates. A large number of network codes have been developed to solve these equations. Figure~\ref{fig:shell-network-comparison} compares 4 different nuclear networks modeling the nucleosynthetic yields of 3 different isotopes for a single shell-burning trajectory using default nuclear physics.  Especially for $^{51}$Cr, the abundance evolution can be very different (varying by nearly an order of magnitude). However, final abundances do not differ by more than a factor of 2. 
%$\sim$10-20\%.  

Varying results can arise from different nuclear inputs (including reaction rates, nuclear masses, partition functions, and weak interaction rates), numerical methods (e.g., differences in the integrators or matrix solvers), but can also arise from the implementation of additional physics (e.g. electron screening).  Here we discuss a few of these uncertainties.

\begin{figure}
    \centering
    \includegraphics[width=\linewidth]{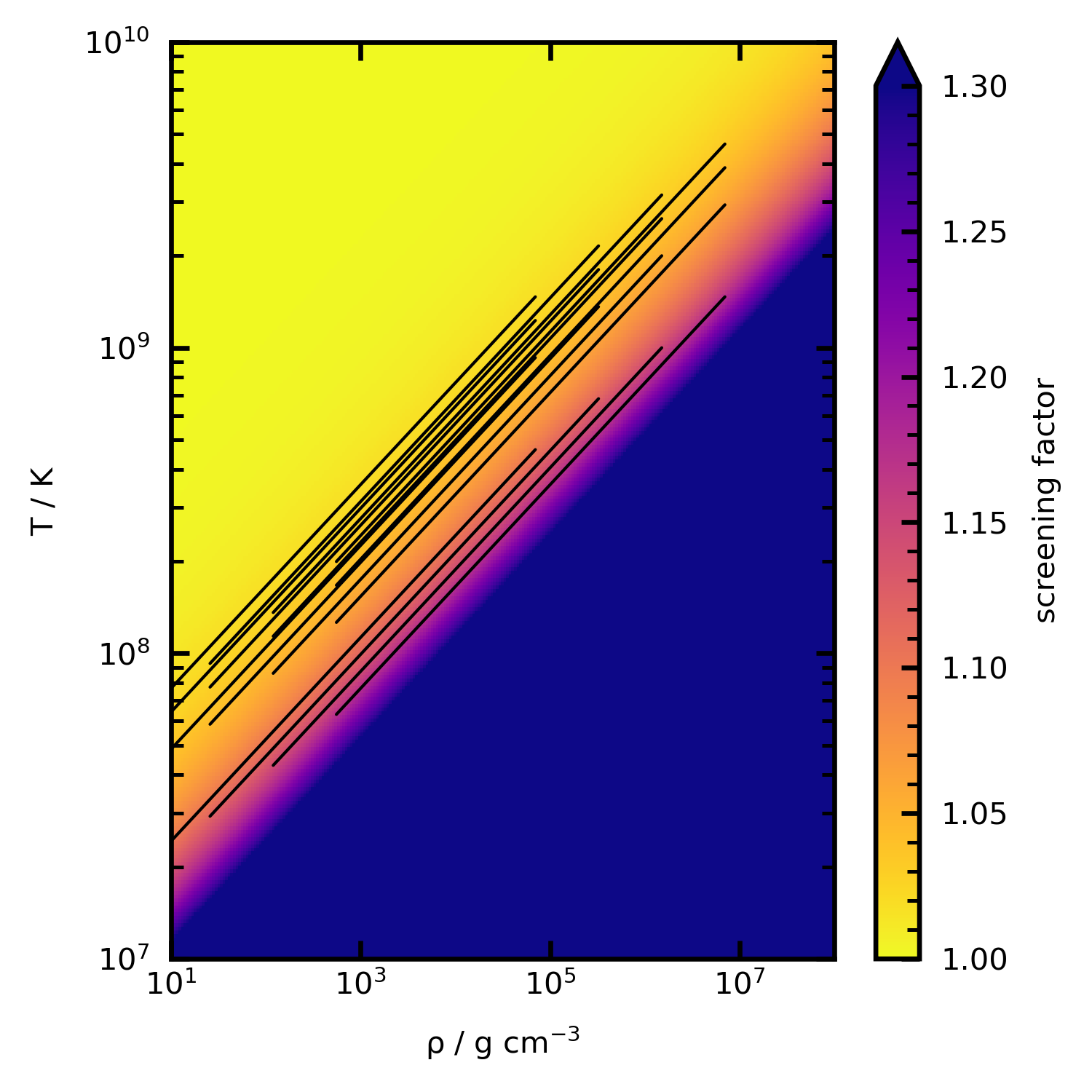}
    \caption{Screening factor for the $^{22}\mathrm{Ne}(\alpha,n)^{25}\mathrm{Mg}$ reaction using the screening prescription from ~\cite{2016ApJS..225...24P}. Shell shock trajectories are plotted as black lines. The shell trajectories are mostly affected by screening in our calculations.}
    \label{fig:shell-screening}
\end{figure}

Any network should also include an implementation of the effect of electron screening on the coulomb barrier for charged-particle reactions.  Figure~\ref{fig:shell-screening} shows the effect of screening on the rates along with a series of shell-burning trajectories using the screening prescription from ~\cite{2016ApJS..225...24P}.  At high temperatures, screening effects are minimal and, for the trajectory studied in Figure~\ref{fig:shell-network-comparison}, the differences caused by varying the screening implementations are minimal.  However, at lower temperatures, differences in the implementation of screening can produce large differences.

A number of uncertainties arise from various approximation choices.  For example, every network must also implement a solution for trajectories that drive the evolution outside of the network bounds.  Many models assume an instantaneous decay that, depending on the conditions, can lead to erroneous results.  This effect can be minimized by increasing the network size. Some networks are adaptive and do this automatically. 

An important question is the significance of numerical uncertainties or other code differences. Identifying such additional uncertainties would require running different networks with exactly the same nuclear physics input and screening prescriptions in order to disentangle input differences from intrinsic code differences. \cite{Lippuner2017} carried out such a study for a range of astrophysical scenarios comparing their SkyNet code with WinNet and Xnet. Overall they found negligible differences between the codes. They did emphasize the importance of calculating reverse rates accurately using the detailed balance principle. 

\subsection{Sensitivity studies}
\label{sec:SensitivityStudies}

It is important to identify the set of critical nuclear reactions that addresses a particular astrophysical question. Beyond the fundamental interest in identifying the nuclear physics effects that governs the universe, this is important to guide future nuclear physics work in experiment and theory.  This critical set of nuclear reactions depends both on the properties of the explosion and the nuclear isotopes being studied.  The various methods used to address this problem fall broadly into two categories: sensitivity studies and error propagation studies. Sensitivity studies answer the question "What are the important reactions in the network that affect the observable". They represent a mathematical analysis of the properties of the reaction network. In the simplest approach all reactions are individually varied one-by-one by the same pre-defined factor. Reactions can also be varied in pairs or larger groups to account for correlations, or the Monte-Carlo technique can be used. The results are usually independent of the actual uncertainties of the reaction rates. Sensitivity studies are important because estimating reaction rate uncertainties is challenging. The reason for this is that most astrophysical charged-particle reaction rates are not measured directly, but inferred from either theory or a complex mixture of direct and indirect experimental studies that are incomplete and complemented with theory. The theory approaches currently used lack a rigorous quantified uncertainty. Sensitivity studies can then be used to identify reactions for which a special effort can be made to estimate uncertainties, and which can be revisited when new information emerges. Furthermore, even if a rate is considered to have a very low uncertainty with no impact on the observable, one may choose to confirm this result with additional measurement if it is a critical reaction. Sensitivity studies provide the guidance for such efforts. If the follow-on question is "What should I measure next" sensitivity study results can be folded with uncertainty estimates to mimic an error propagation \citep[e.g.,][]{Hermansen2020}. However, it is difficult to account for any correlations in uncertainties in that approach. 

Error propagation analysis answers the question "What are the dominant nuclear uncertainties in the prediction of an observable". They are based on assumptions on the uncertainties of nuclear reaction rates, and propagate these uncertainties to the final observable. We discuss this type of study further in Section \ref{sec:NucErrorPropagation}. Error propagation methods are critical for quantitative comparison  of models with observations, and are also important to guide nuclear experimental efforts that seek to address the most important nuclear uncertainties. They can, in principle, also be used as sensitivity studies by looking for correlations between input and output variations. However, it can be challenging to identify important reactions whose uncertainty estimates are incorrect, or in scenarios where a large number of comparable uncertainties contributes to the observable. 

A number of nucleosynthesis sensitivity studies related to core collapse supernovae have already been performed in which reaction rates are varied (one-by-one), and their impact on key nuclides is quantified. A summary of their results for the radioisotope, $^{44}$Ti, is shown in Tab.~\ref{tab:44Ti-Sensitivity}. Only the reactions whose uncertainties are most impactful are shown for the sake of brevity. The column labeled ``Sensitivity'' is a measure of how much the abundance of $^{44}$Ti changes when that reaction is varied within the uncertainties tabulated in the \texttt{STARLIB} library. This method was already employed by \cite{Hermansen2020} so we adopt their sensitivities directly. For the findings of \cite{Subedi2020} and \cite{Magkotsios2010}, the reactions were varied with a fixed factor, so the default \texttt{STARLIB} reaction rate uncertainty for each reaction was used to scale those sensitivities appropriately.

In the present work, we performed an additional sensitivity study using the trajectory described in Eq.\ref{eq:bradsTrajectory}, in which every reaction was varied one-by-one as described above. Here, we varied each reaction within its assumed uncertainty in the \texttt{Starlib}. A key feature of the library is that the reaction rates are tabulated along with their temperature-dependent uncertainties. It is important to consider the reaction's uncertainty at the temperature at which the nuclear burning is occurring. The techniques described in Sec.~\ref{sec:NucErrorPropagation} are adopted to achieve this and the results are included in Tab.~\ref{tab:44Ti-Sensitivity}. Once again, the $^{44}$Ti($\alpha$,p)$^{47}$V reaction is identified as a key reaction for this model because the default \texttt{starlib} uncertainty is a factor of 10. On the other hand, the titanium-producing $^{40}$Ca($\alpha$,$\gamma$)$^{44}$Ti reaction is not highlighted because its uncertainty at the burning temperatures is well constrained experimentally. In addition to this, the $^{27}$Al($\alpha$,n)$^{30}$P and $^{39}$K(p,$\alpha$)$^{36}$Ar reactions are identified as important reactions, whose uncertainties affect the $^{44}$Ti yield by almost a factor of 2 when varied within their uncertainties. At present, their rates in the \texttt{Starlib} library are from TALYS with a large adopted factor-of-10 uncertainty.

There are a number of key take-aways from this table: (i) The vast majority of the reactions governing $^{44}$Ti production in supernovae remain unmeasured; (ii) some reactions, although measured, are not included in the default \texttt{Starlib} database.

\begin{table*}
  \centering
  \begin{tabular}{l|lcccll}
    \hline \hline
    Status                                      & Reaction                              & Sensitivity & \texttt{STARLIB} & \texttt{STARLIB} & Reaclib & Target     \\
                                                &                                       &             & source           & factor unc.      & source  & stability  \\ \hline
    \multirow{2}{*}{Some experimental data} & $^{44}$Ti($\alpha$,p)$^{47}$V         & 2.61        & TALYS            & 10               & chw0    & long lived \\
                                                & $^{40}$Ca($\alpha$,$\gamma$)$^{44}$Ti & 1.14        & po13             & 1.07 -- 1.76     & chw0    & stable     \\ \hline
    \multirow{6}{*}{Limited or no experimental data}              & $^{27}$Al($\alpha$,n)$^{30}$P         & 1.7         & TALYS            & 10               & HF      & stable     \\ 
                                                & $^{39}$K(p,$\alpha$)$^{36}$Ar         & 1.4         & TALYS            & 10               & HF      & stable     \\
                                                & $^{43}$Sc(p,$\alpha$)$^{40}$Ca        & 1.15        & TALYS            & 10               & HF      & unstable   \\
                                                & $^{43}$Sc(p,$\gamma$)$^{44}$Ti        & 1.18        & TALYS            & 10               & HF      & unstable   \\
                                                & $^{45}$V (p,$\gamma$)$^{46}$Cr        & 1.1 -- 2.0  & TALYS            & 10               & HF      & unstable   \\
                                                & $^{17}$F ($\alpha$,p)$^{20}$Ne        & 1.2 -- 1.3  & TALYS            & 10               & NACRE   & unstable   \\
                                                & $^{57}$Co(p,$\gamma$)$^{58}$Ni        & 1.2         & TALYS            & 10               & HF      & unstable   \\ \hline \hline
  \end{tabular}
  \caption{Reaction cross sections identified as being important for
    $^{44}$Ti production in supernova explosions. These are compiled
    from \cite{Hermansen2020,Subedi2020,Magkotsios2010}.}
  \label{tab:44Ti-Sensitivity}
\end{table*}

\subsection{Propagating Nuclear Uncertainties to Observable Yields}
\label{sec:NucErrorPropagation}

Investigating the impact of nuclear physics uncertainties on the
production of $\gamma$-ray emitters requires varying reaction rates
within their uncertainties to ascertain their impact on nuclide
production in the stellar model. There are different approaches of how this can be accomplished. In the simplest approach one varies all reactions independently within their estimated uncertainties using a Monte-Carlo method (e.g. \cite{Nishimura:2019,Denissenkov2018} for other astrophysical scenarios). This approach neglects correlations between uncertainties that emerge primarily from the use of the same theoretical model for a relatively large number of reactions that have no experimental constraints. An alternative approach that can address this issue to some extent is a Monte Carlo method, where for theoretical rates one varies parameters of the theoretical ingredients, which are then applied to all predicted reaction rates (e.g. \cite{Goriely2021,Bertolli2013} for other astrophysical scenarios). The former approach has the advantage to provide still some information of individual rate sensitivities and can thus provide guidance for experimental efforts to measure reaction rates. The latter may provide a more realistic error budget, but uncertainties become strongly dependent on the chosen model framework, and additional uncertainty components may be missed. Guidance for experimental efforts is largely limited to which of the theory ingredients need to be constrained experimentally. There are recent efforts to combine the two methods to explore statistical and correlated uncertainties in a unified framework (\cite{Martinet2024}). 

Here we explore the impact of our estimated reaction rate uncertainties on the production of $^{44}$Ti using the simple independent Monte Carlo approach. 
While this is, in principle, a
trivial endeavor, Tab.~\ref{tab:44Ti-Sensitivity} shows that each
reaction can have very different uncertainties. Furthermore, some
reactions, particularly those constrained by experimental information,
have temperature-dependent uncertainties. For the purposes of this
study, we leverage the \texttt{STARLIB} reaction rate library
described in \cite{Sallaska2013}\footnote{Available for download
  from \url{https://github.com/Starlib/Rate-Library}}. 

The default \texttt{STARLIB} reaction rate library currently does not
include all the most recent experimental constraints on reaction
rates. It also assumes that any reaction rates based on the
Hauser-Feshbach statistical model have a factor of 10 uncertainty.
Section \ref{sec:HFVariations} outlines the process that should be
undertaken in future work to better characterize these uncertainties for the
reactions that create $^{44}$ Ti.

The \texttt{STARLIB} reaction rate library is a \textit{tabulated}
library of reaction rates as a function of temperature. The distinct
advantage of using tables in this case is that the uncertainty of the
rate as a function of temperature can be trivially added as a third column. As
pointed out in \cite{Longland2010}, there are both physical and
statistical reasons to believe that the probability distribution of a
reaction rate is distributed lognormally. As such, it is possible to
describe the full probability distribution of the rate at a given
temperature by just two numbers: 'lognormal $\mu$' and 'lognormal
$\sigma$'. The distribution, $f(x)$ is given by:
\begin{equation}
  \label{eq:lognormal}
  f(x) = \frac{1}{\sigma\sqrt{2\pi}} \frac{1}{x}e^{-(\ln x - \mu)^2/(2\sigma^2)},
\end{equation}
where $x$ represents the reaction rate in this case. These parameters
are also naturally translated into colloquial descriptions of reaction
rates: the ``recommended rate'' and the ``factor uncertainty'',
$f.u.$:
\begin{equation}
  \label{eq:FU}
  \text{Recommended Rate} = e^{\mu}, \qquad f.u. = e^{\sigma}.
\end{equation}
The recommended rate and factor uncertainty are tabulated in
\texttt{STARLIB}. In the simplest case, these parameters can be found
from the recommended ('rec.'), ``high'', and ``low'' rates by:
\begin{equation}
  \label{eq:musig}
  \mu = \ln x_{\text{rec.}}, \qquad \sigma = \ln \sqrt{\frac{x_{\text{high}}}{x_{\text{low}}}}.
\end{equation}
For more details on how these parameters can be obtained using a more
thorough statistical process, the interested reader is referred to
\cite{Longland2010} and \cite{Sallaska2013}.

Once a reaction rate library contains temperature-dependent rates and
uncertainties, the question becomes one of how to use it. For example,
how does one draw a random sample of a reaction rate if the
uncertainty is a temperature-dependent quantity? It is not always
appropriate to multiply and divide a rate by a factor of 100, for
example. \cite{Longland2012b} investigated this problem and
proposed a computationally straight-forward method for achieving this.
The method relies on the fact that the rate probability distributions
are distributed lognormally. Given this, a single sample of the
reaction rate at temperature $T$ can be represented by
\begin{equation}
  \label{eq:RRSample}
  x(T) = e^{\mu(T)}\cdot e^{p(T)\sigma(T)} = x(T)_{\text{rec.}} (f.u.)^{p(T)}.
\end{equation}
The parameter $p(T)$ is referred to as the rate variation parameter and is a normally-distributed random variable (i.e. it has a mean of 0 and standard deviation of 1). \cite{Longland2012b} found that it is sufficient to treat it as constant with respect to temperature. To visualize this, consider a case where the factor uncertainty, $f.u.$, is 10 for a reaction at all temperatures. To calculate nucleosynthesis at the ``high'' rate (a factor of 10 above the recommended rate), one would use $p=1$. For an experimentally-constrained reaction rate with temperature-dependent uncertainties, the ``low'' rate at \textit{all} temperatures can be found by using $p=-1$. This tool makes Monte Carlo nucleosynthesis studies over a wide range of reactions with varying uncertainties trivial.

\begin{figure}[!ht]
  \centering
  \includegraphics[width=0.5\textwidth]{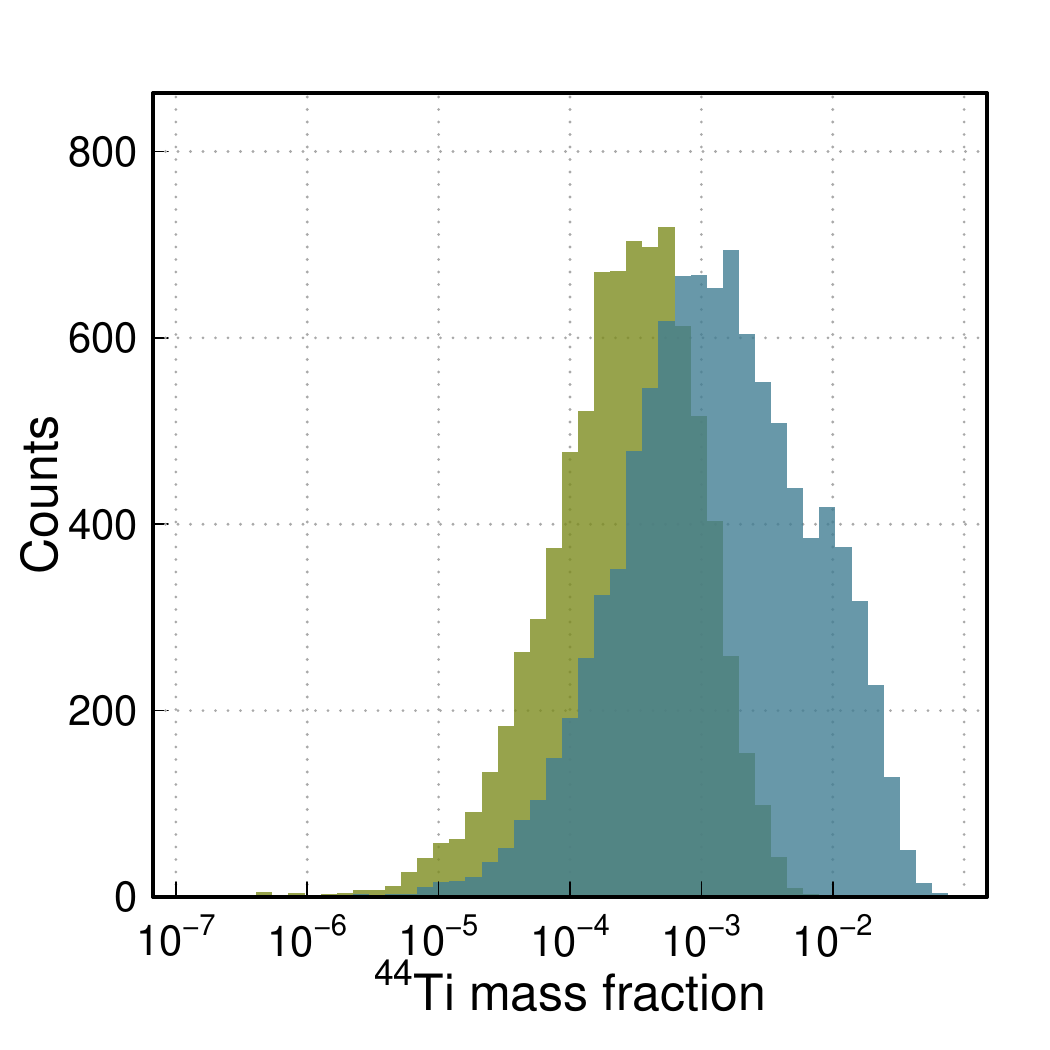}
  \caption{Variation in $^{44}$Ti production when nuclear reaction
    rates are varied within their default \texttt{STARLIB}
    uncertainties. In green, the $^{44}$Ti abundance resulting from an
    adiabatic expansion are shown, whereas the blue depicts abundances
    resulting from the base trajectory in Section~\ref{sec:NucErrorPropagation}.}
  \label{fig:44Ti-MC}
\end{figure}

Using the default \texttt{STARLIB} reaction rate library, 10,000 reaction network calculations were performed on both the adiabatic trajectory given by Eqs.~\eqref{eq:adiabatic1} and \eqref{eq:adiabatic2} and a trajectory using equation~\ref{eq:bradsTrajectory} with $a=10.0$, $\tau_1=0.1$, $b=0.5$, $n=1.0$, $t_0=3.0$, $\tau_2=0.3$, and $\rho_0=10^7 {\rm g \, cm^{-3}}$.  This trajectory was found to produce $^{44}$Ti in the central region of the supernova explosion. This is our base trajectory.  In Section~\ref{sec:Ti44}, we will vary $b$ in this base trajectory to study the role of reheating.  For each realization of the nucleosynthesis, every reaction rate in the network was varied simultaneously within its given uncertainty. The results of these calculation are shown in Fig. \ref{fig:44Ti-MC}. 

Clearly there is a very large uncertainty in $^{44}$Ti production just from considering the nuclear physics uncertainties alone. Indeed, the nuclear physics uncertainties alone are larger than the change in predicted abundance from the two trajectories considered here. The predicted $^{44}$Ti mass fraction for the adiabatic expansion model is $X=1.3 \times 10^{-4}$ with a factor of 4.3 uncertainty, whereas the prediction from the base trajectory (Section~\ref{sec:NucErrorPropagation}) is $X=1.3 \times 10^{-3}$ with a factor of 5.0 uncertainty.

\section{Focused Examples in Core-Collapse Supernovae}
\label{sec:examples}

Here we discuss the combined uncertainties on 2 different focused supernova yield studies.

\subsection{Example:  Radioactive Isotopes in the Inner Ejecta}
\label{sec:Ti44}

The most direct probe of the central engine is to study the elements produced in the engine itself.  The central engine reaches such high temperatures that the material is in nuclear statistical equilibrium (NSE) and, for material with $Y_e\approx0.5$ (equal amounts of neutron and protons, $n_n=n_p=n_e$), the dominant isotope produced is $^{56}$Ni.  The early emergence of $\gamma$-rays from $^{56}$Ni decay~\citep{1988ApJ...329..820P} is what led to the development of the current convection-enhanced paradigm for core-collapse supernovae~\citep{1994ApJ...435..339H}.  Studying the Doppler shift of the $\gamma$-ray lines produced from $^{56}$Ni decay provided a first estimate of the nature of the asymmetries in supernovae~\citep{2005ApJ...635..487H}.  $^{44}$Ti is also produced in the ejecta processed in the inner engine in nearly the same inner conditions that produce $^{56}$Ni.  $^{44}$Ti is much more sensitive to the exact conditions of the explosion providing complementary constraints to the $^{56}$Ni.  It is an ideal probe of the exact nature of the explosion (if we can remove the nuclear physics uncertainties).  The $^{44}$Ti observations by NuSTAR remain the strongest proof of the convective engine for supernovae~\citep{2014Natur.506..339G,2017ApJ...834...19G}.  $^{48}$V and $^{48}$Cr are also produced in this inner engine and, for a nearby supernova, could provide complementary constraints~\citep{2020ApJ...890...35A}.

\begin{figure}
  \centering
  \includegraphics[width=0.4\textwidth]{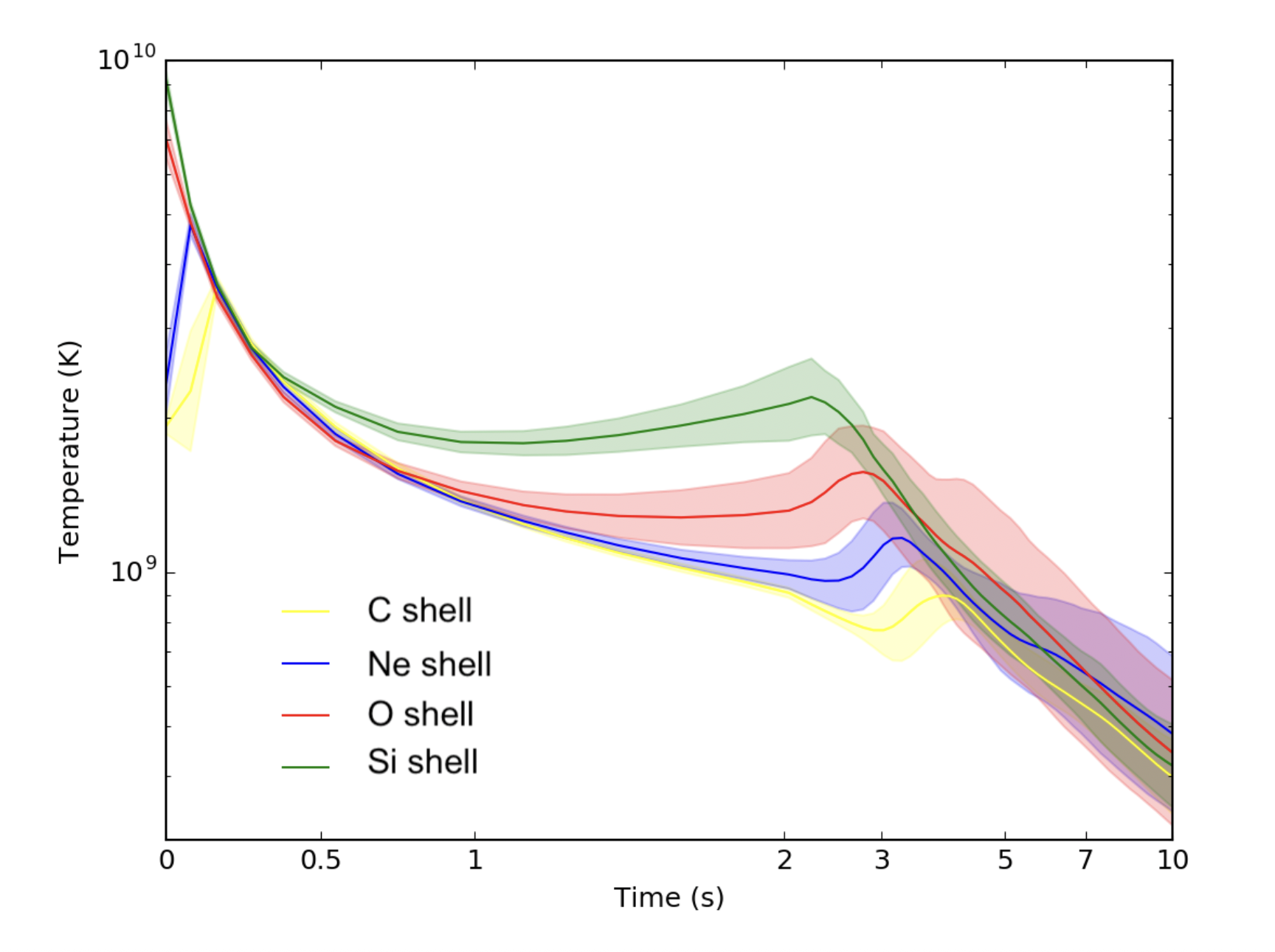}
  \caption{Temperature versus time for ejecta in 3-dimensional supernova calculations~\citep{2020ApJ...895...82V} showing the reheating that occurs as the shock decelerates as it propagates through different burning layers (recall Figure~\ref{fig:dec-star}).}
  \label{fig:vancetraj}
\end{figure}

$^{44}$Ti is subject to all of the uncertainties in both the explosion and nuclear physics discussed above.  Here we probe a few of those uncertainties in detail.  One example of this is shock reheating.  In Section\ref{sec:astro}, we discussed a number of mechanisms (e.g. shock deceleration, continuously-driven engine) that would cause the ejecta to reheat as it propagates through the star.  Figure~\ref{fig:vancetraj} shows one such example of this reheating from a 3-dimensional explosion calculation~\citep{2020ApJ...895...82V}.  In this model, shock deceleration as the shock propagates through the convective regions of the star (see Figure~\ref{fig:dec-star}), it reheats.  

%\textcolor{red}{The functional form for trajectories in Fig.~\ref{fig:Trajectories}. These are from Brad's analytic form for SNe inner layers:
%Here, the first term and the second term parameterize the initial drop in temperature and the subsequent rise, respectively. $a$ is the starting temperature, $\tau_1$ is the timescale of first peak, $b$ is the second peak height, $t_0$ is the position of the second peak, $\tau_2$ is the timescale of decline of the second peak, i.e. corresponds to the width of the second peak, and $t^n$ parameterizes the slope of rise of the increase.
%}
% full functional form : T_9(t) = a * \exp(-t/\tau_1) + b * t^n * (1 + \tanh((t - t_0)/\tau_2) - b * \delta_{n,0} * (1 + \tanh((t_0/\tau_2)) , but n is not set to 0, so the last piece is irrelevant.

\begin{figure}
  \centering
  \includegraphics[width=0.4\textwidth]{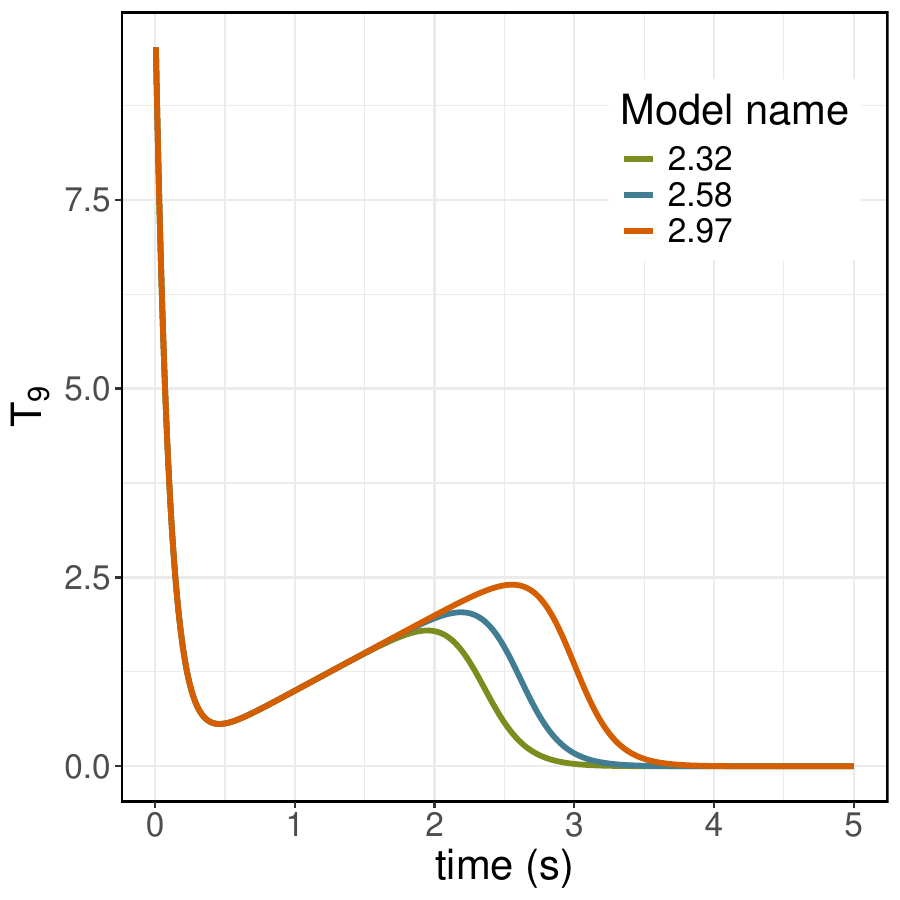}
  \caption{Three trajectories using the equation~\ref{eq:bradsTrajectory} with the same starting temperature but varying the strength of reheating:  $b=$ 2.32, 2.58, and 2.97.  For these three values, the reheating leads to second peak temperatures from $1.75\times 10^9$K to $2.5 \times 10^9$K.}
  \label{fig:Trajectories}
\end{figure}

Using equation~\ref{eq:bradsTrajectory} we create three different values of $b$ that describes the magnitude of the reheating:  Figure~\ref{fig:Trajectories}.  Using these trajectories to model the material ejected from the inner-engine-region, the $^{44}$Ti production yield can be further investigated in the presence of nuclear physics uncertainties. Following the same strategy as laid out in Sec.~\ref{sec:NucErrorPropagation}, a Monte Carlo method is used to propagate the nuclear reaction rates to the final $^{44}$Ti yield for trajectories $b=$2.32, 2.58, and 2.97.  Two prominent observations can be made from the figure: (i) the $^{44}$Ti yield is very sensitive to the exact conditions in the re-heated trajectory (this concept was also investigated in great detail for traditional trajectories in \cite{Magkotsios2010}); and (ii) the nuclear physics uncertainties affect the yield at roughly the same level as the thermodynamic conditions. For example, the Monte Carlo $^{44}$Ti yields obtained from the 2.58 model span mass fractions $X(^{44}$Ti$) = 0$ -- $0.05$, which is a larger variation than single nucleosynthesis calculations for these trajectories would imply.

\begin{figure}
  \centering
  \includegraphics[width=0.4\textwidth]{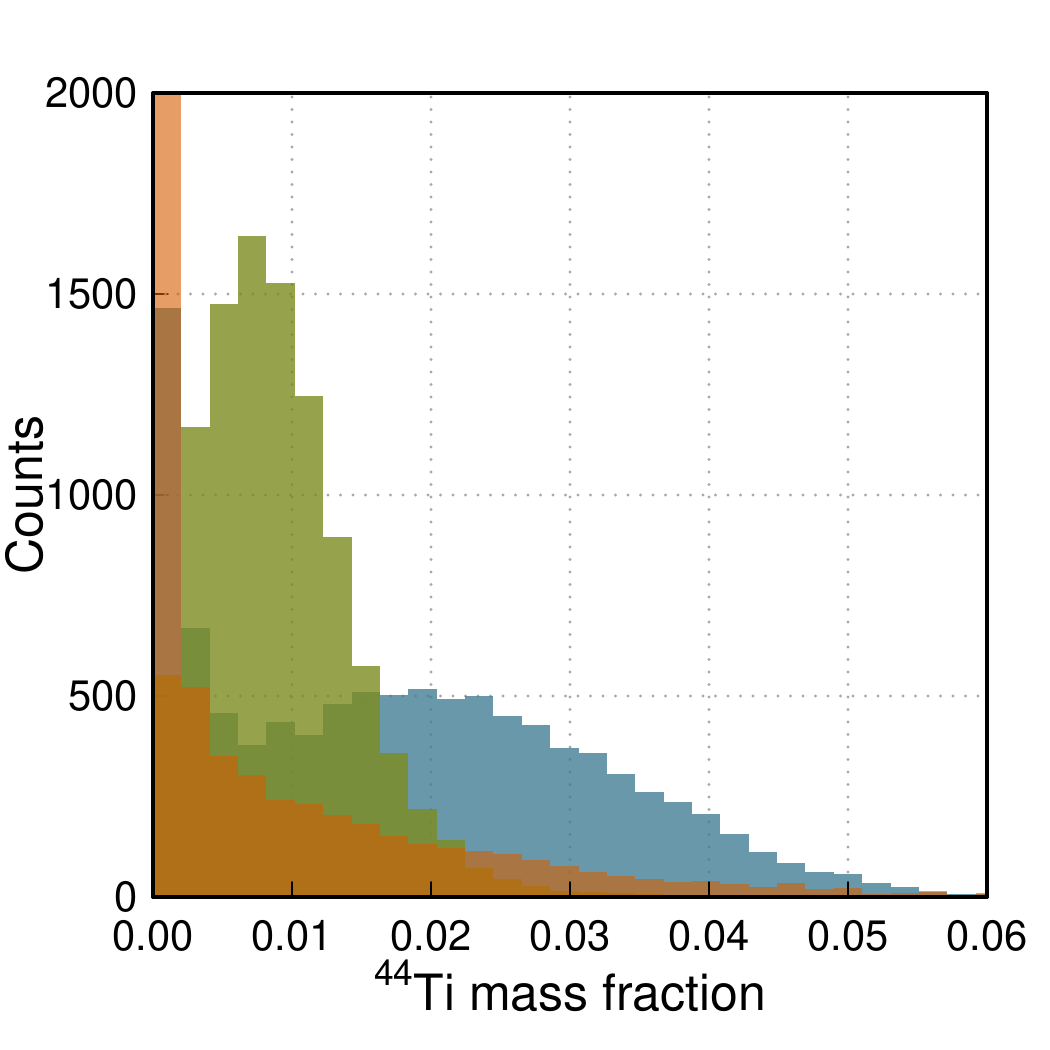}
  \caption{$^{44}$Ti yields from the trajectories shown in Fig. \ref{fig:Trajectories}.  For each trajectory, we include the variation caused by nuclear physics uncertainties (recall Figure~\ref{fig:44Ti-MC}).  These mild variations in the trajectories can cause large shifts in the $^{44}$Ti yield.}
  \label{fig:44Ti-MC-NewTrajectories}
\end{figure}

\begin{figure}
    \centering
    \includegraphics[width=\linewidth]{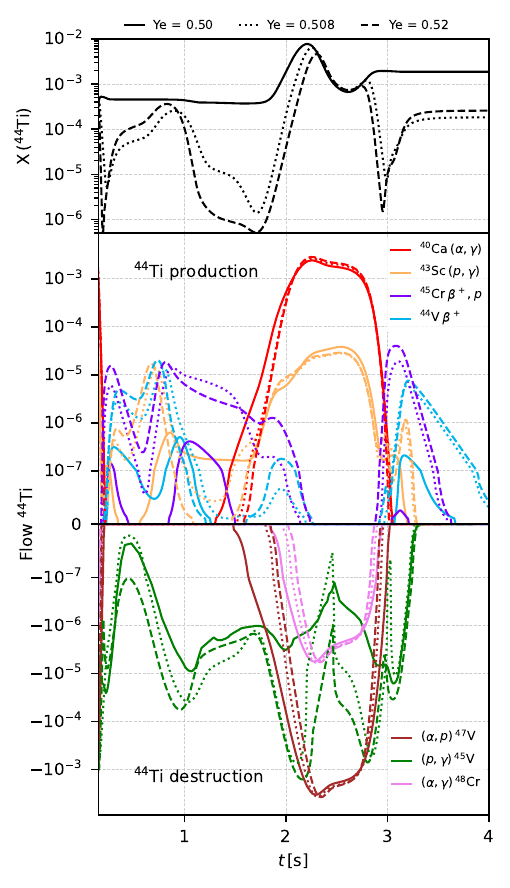}
    \caption{Mass fraction of $^{44}$Ti (top), production flow (middle) and destruction flows (bottom) as a function of time for the same trajectory evolution (using our base trajectory - Section~\ref{sec:NucErrorPropagation}) but with electron fractions $Y_e=0.50,0.508,$ and $0.52$ (solid, dotted and dashed, respectively). Network code used: WinNet.}
    \label{fig:ti44_flow_ye}
\end{figure}

Another uncertainty is due to the exact value of the electron fraction. $^{56}$Ni is only mildly sensitive to the exact electron fraction, whereas $^{44}$Ti is extremely sensitive to this value.  Increasing the value of the electron fraction above 0.5 by a couple per cent can increase the destruction rate of $^{44}$Ti, leading to an order of magnitude decrease in the final $^{44}$Ti yield (Fig.~\ref{fig:ti44_flow_ye}). The main production and destruction channel remains the same in all three conditions, $^{40}$Ca($\alpha,\gamma$)$^{44}$Ti and $^{44}$Ti($\alpha,p$)$^{47}$V, however, in proton rich conditions, new reaction channels on the $p>n$ side start contributing to the inwards and outwards flow. In particular, at about 3.1s $^{45}$Cr$\rightarrow^{44}$Ti + $e^+ + p + \nu_e$ opens a new pathway to produce $^{44}$Ti, while new destruction mechanisms via proton absorption also open up. $^{45}$Cr has a 34\% probability to emit a proton along with a $e^+$-decay (Nudat 3.0 from the National Nuclear Data Center~\url{https://www.nndc.bnl.gov/nudat3/}) leading to the beta+-decay-delayed proton emission to produce $^{44}$Ti. This pathway requires substantial amounts of free protons and is opened especially at $Y_e = 0.52$ causing a net increase in $^{44}$Ti abundance in this compared to $Y_e = 0.508$. 

As can be seen in Figure~\ref{fig:ti44_flow_ye}, the production and destruction of $^{44}$Ti depends upon a number of rates.  As we showed in Figure~\ref{fig:44Ti-MC}, these uncertainties alone can lead to order of magnitude errors in the $^{44}$Ti production.  To use $^{44}$Ti to probe the nature of the supernova explosion, we ultimately need to reduce these nuclear physics uncertainties.

%\begin{figure}
%    \centering
%    \includegraphics[width=\linewidth]{compare_nn_inner.pdf}
%    \caption{Relative error for the predicted production of $^{44}$Ti and $^{48}$Cr between different nuclear networks for the set of analytic trajectories described in the main text [Brad's trajectories]}
%    \label{fig:compare-nn-inner}
%\end{figure}

We have already listed examples of how $\gamma$-ray observations from both $^{56}$Ni and $^{44}$Ti have helped solidify the current paradigm behind supernova explosions.  The COSI instrument~\citep{2024icrc.confE.745T,tomsick2019comptonspectrometerimager} will provide additional information for both the Cassioepeia A remnant and SN 1987A, bolstering the NuSTAR results~\citep{2014Natur.506..339G,2015Sci...348..670B,2017ApJ...834...19G}.  Improved energy resolution in our observations allow us to use the Doppler broadening in the $^{44}$Ti lines to both confirm the structure inferred from the $^{56}$Ni decay-line measurements and, by leveraging the sensitivity of $^{44}$Ti production, further probe details of the explosive engine.  

\begin{figure}
  \centering
  \includegraphics[width=0.4\textwidth]{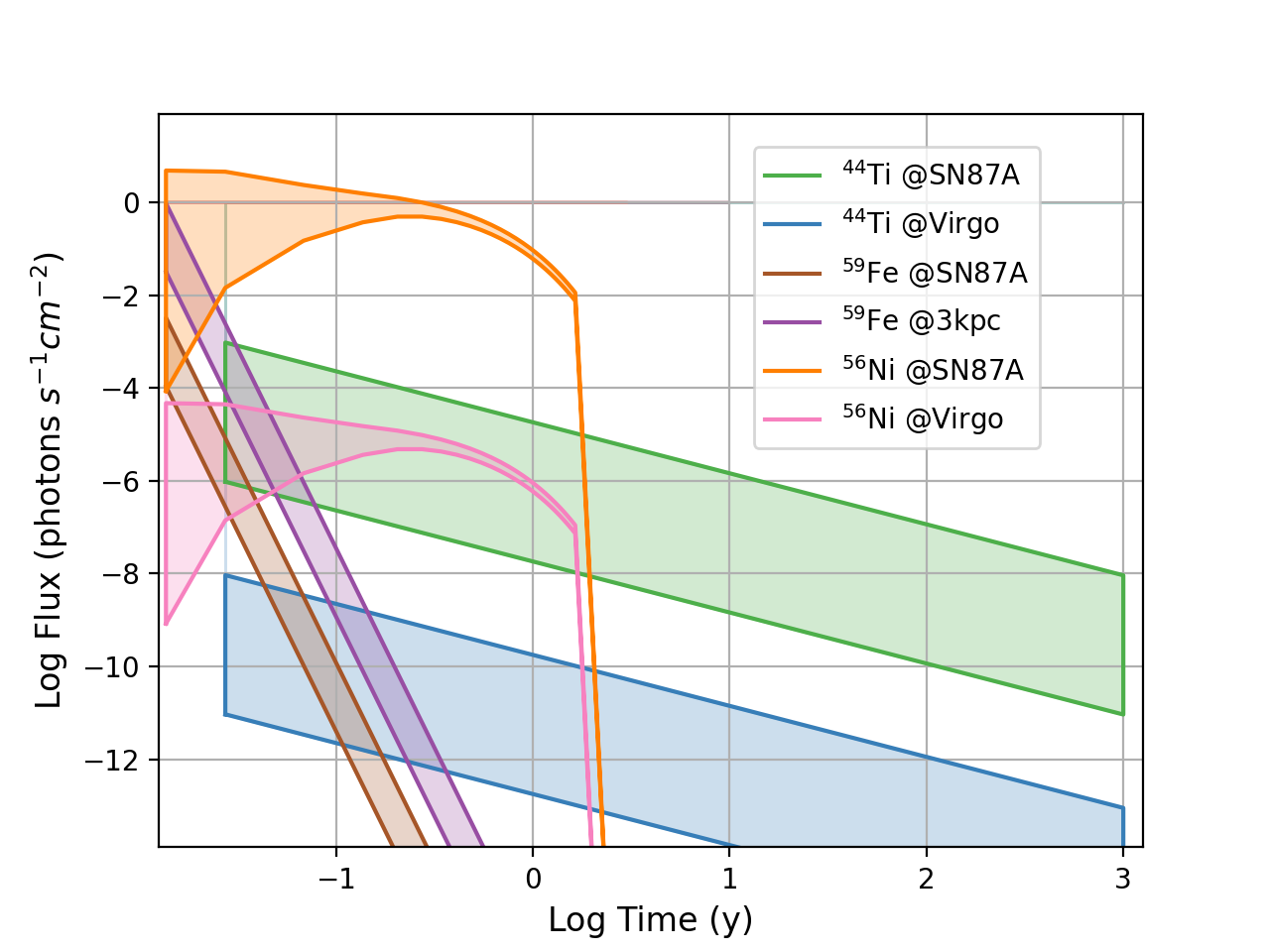}
  \caption{Gamma-ray fluxes as a function of time since the explosion for both $^{56}$Ni and $^{44}$Ti.  These results show the expected fluxes for an explosion at the distance of SN 1987A and for supernovae occuring in the Virgo cluster.  The range in the $^{56}$Ni signal is primarily due to the amount of mixing (and, to a lesser extent), the strength of the explosion~\citep{2023ApJ...956...19F}.   The range in the $^{44}$Ti is primarily due to uncertainties in the nuclear reactions and the trajectories.}
  \label{fig:gammaray}
\end{figure}

Potential future detectors could observe $\gamma$-rays from supernovae out to the Virgo cluster, dramatically increasing the number of observed events.  Figure~\ref{fig:gammaray} shows the expected $^{56}$Ni and $^{44}$Ti signals from supernovae and their uncertainties as a function of observing time.  Upcoming (e.g. COSI) detectors are at the edge of detecting $\gamma$-rays out to the Virgo cluster and next-generation detectors have the potential to detect all supernovae in $^{56}$Ni out to the Virgo Clsuter.  $^{44}$Ti is more difficult to detect, but detailed observations of $^{44}$Ti from SN 1987A will be done by upcoming missions.  Combined with the already existing $^{56}$Ni observations of SN 1987A, scientists will be able to probe the details of the supernova engine.

\subsection{Example:  Shell Burning}
\label{sec:shell}

\cite{2020ApJ...890...35A} identified a number of radioactive isotopes that are predominantly produced when the supernova blast wave propagates through the carbon or helium shells in the star. The dominant production mechanism is as follows. Heating by the shock wave in the carbon shell releases $\alpha$-particles via $(\gamma,\alpha)$ reactions, which produce neutrons via $^{22}\mathrm{Ne}(\alpha,n)$ reactions. In the helium shell, there is also unburned helium that can contribute to the $\alpha$-particles when the shock arrives. The neutrons capture on isotopes of heavier elements either produced in pre-supernova $s$-process or present from the initial composition, creating the radioactive species \citep[see, e.g.][]{jones2019a}.

The carbon and helium shells are sufficiently far out that they are not significantly affected by many late-time engine-activity mechanisms, but they still naturally contain signatures of (1) the stellar structure at collapse and (2) multi-dimensional shock structure. Therefore, measurements of $\gamma$-radiation from isotopes produced in these shells could be used to infer details about the pre-supernova structure of the star.
Of particular interest, owing to their radioactivity and the abundance with which they are produced in this process, are the isotopes $^{22}$Na, $^{24}$Na, $^{65}$Zn, $^{51}$Cr and $^{60}$Fe.

\begin{figure*}
    \centering
    \includegraphics[width=\linewidth]{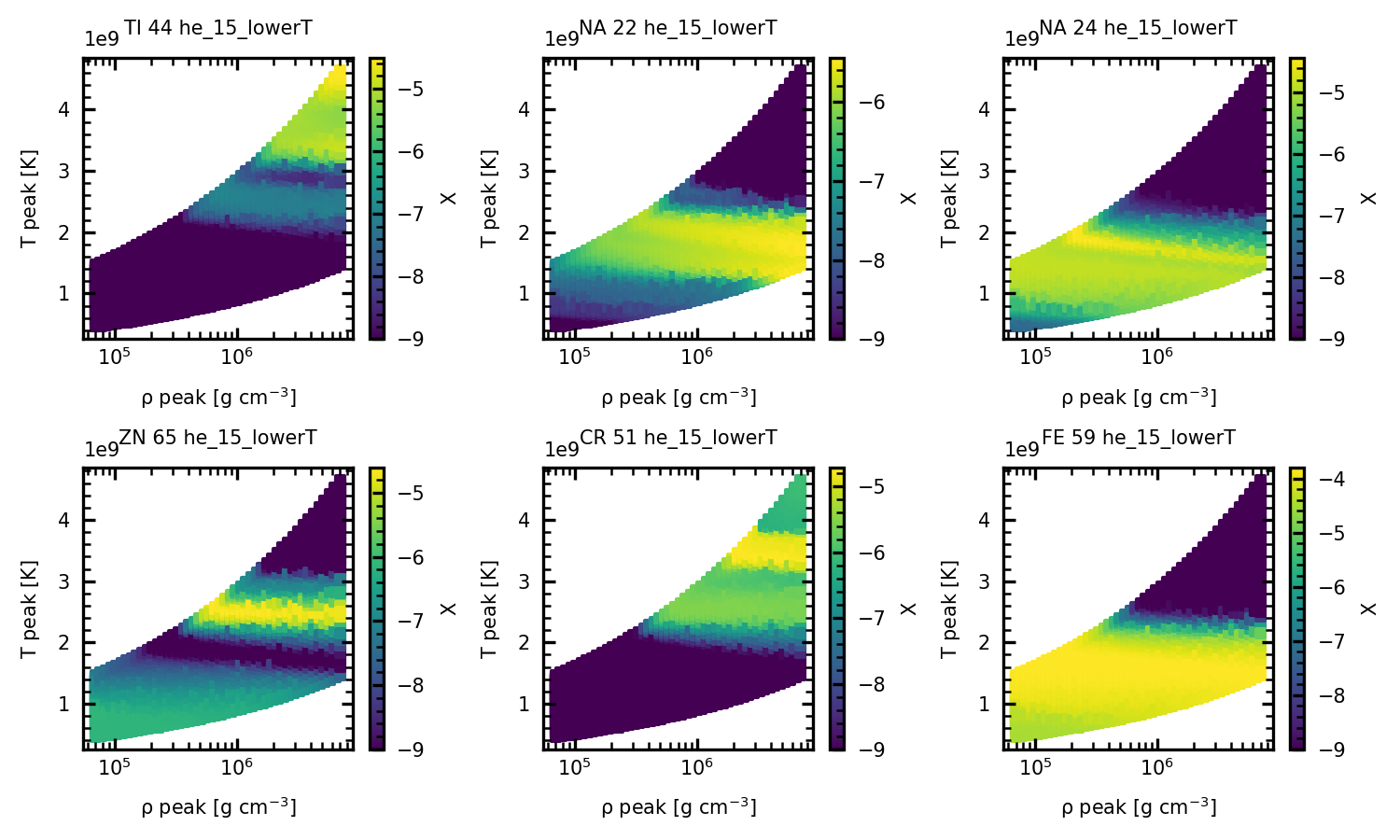}
    \includegraphics[width=\linewidth]{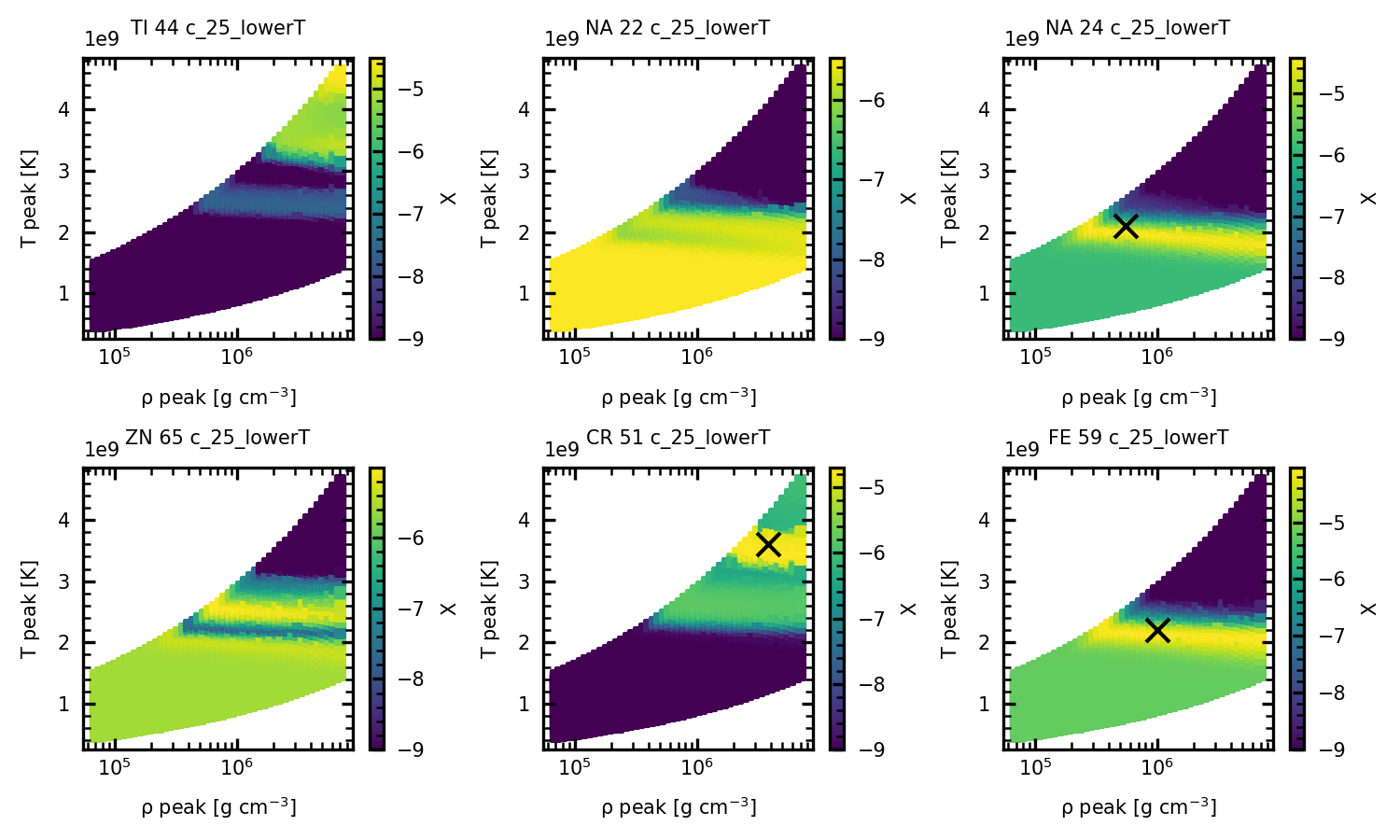}
    \caption{Shock nucleosynthesis of radioactive isotopes in the helium shell of a 15~$M_\odot$~stellar model (top six panels) and the carbon shell of a 25~$M_\odot$~stellar model (bottom six panels).}
    \label{fig:shell-nuc}
\end{figure*}

To get an overview of the conditions favoring the production of these isotopes, we have taken initial compositions characteristic of pre-supernova helium and carbon shells from stellar models with full nucleosynthesis post-processing data \citep{2020ApJ...890...35A}, and approximated their shock nucleosynthesis using a parameterized strong shock solution and exponential decay for the thermodynamic post-shock trajectories. Figure~\ref{fig:shell-nuc} shows the results of this study, where the yield of the isotope is drawn as a color map with respect to the peak density and temperature conditions achieved by the incident. The top six panels were using the helium shell composition from a 15~\msun~model for the pre-shock state and the bottom six panels were using the carbon shell composition from a 25~\msun~model. The production bands are much more sensitive to the post-shock temperature than the density, as could be expected owing to the strong temperature dependence of the reaction rates. The peak production temperatures for the isotopes span quite a range in temperatures, making them potentially useful probes of the conditions reached in these layers of the star during the explosion.

\begin{figure*}
    \centering
    \includegraphics[width=0.33\linewidth]{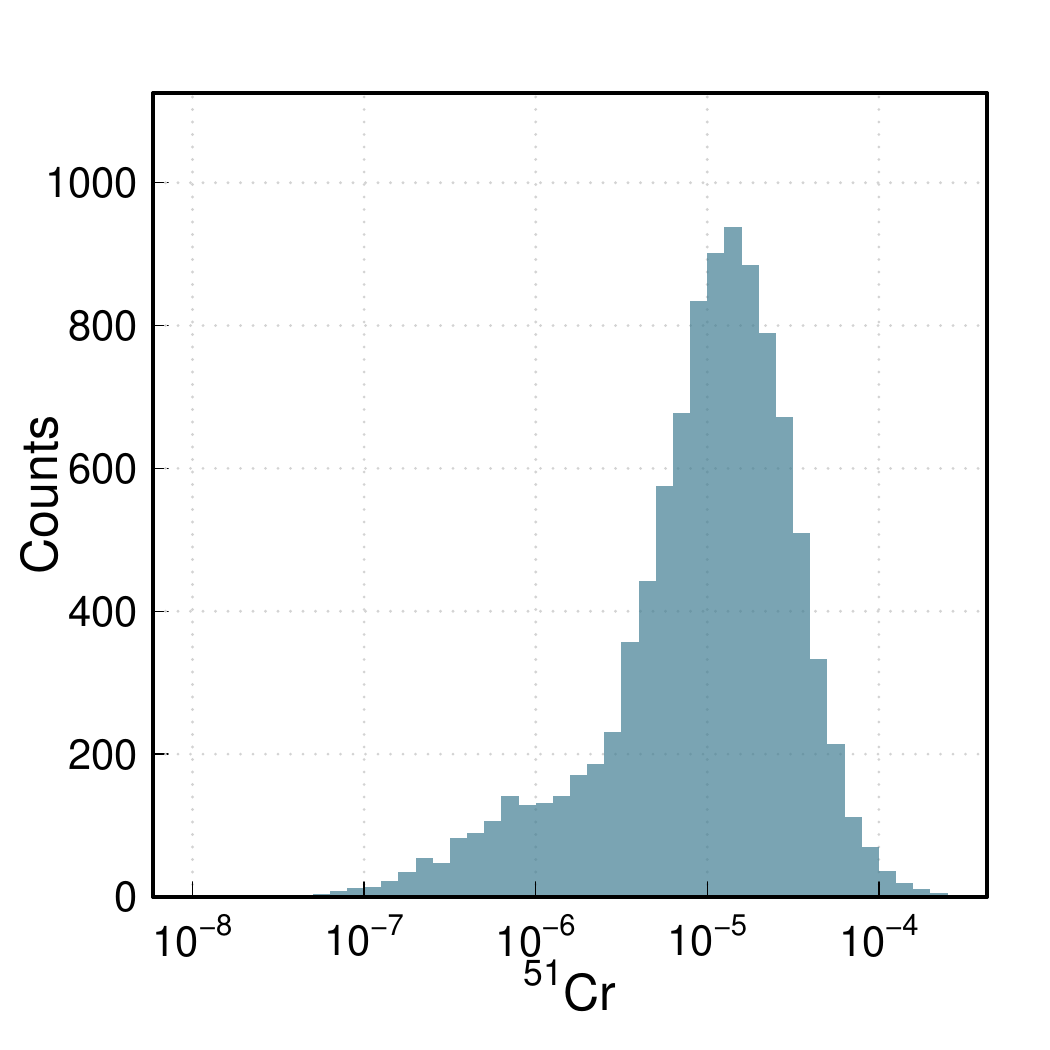}
    \includegraphics[width=0.33\linewidth]{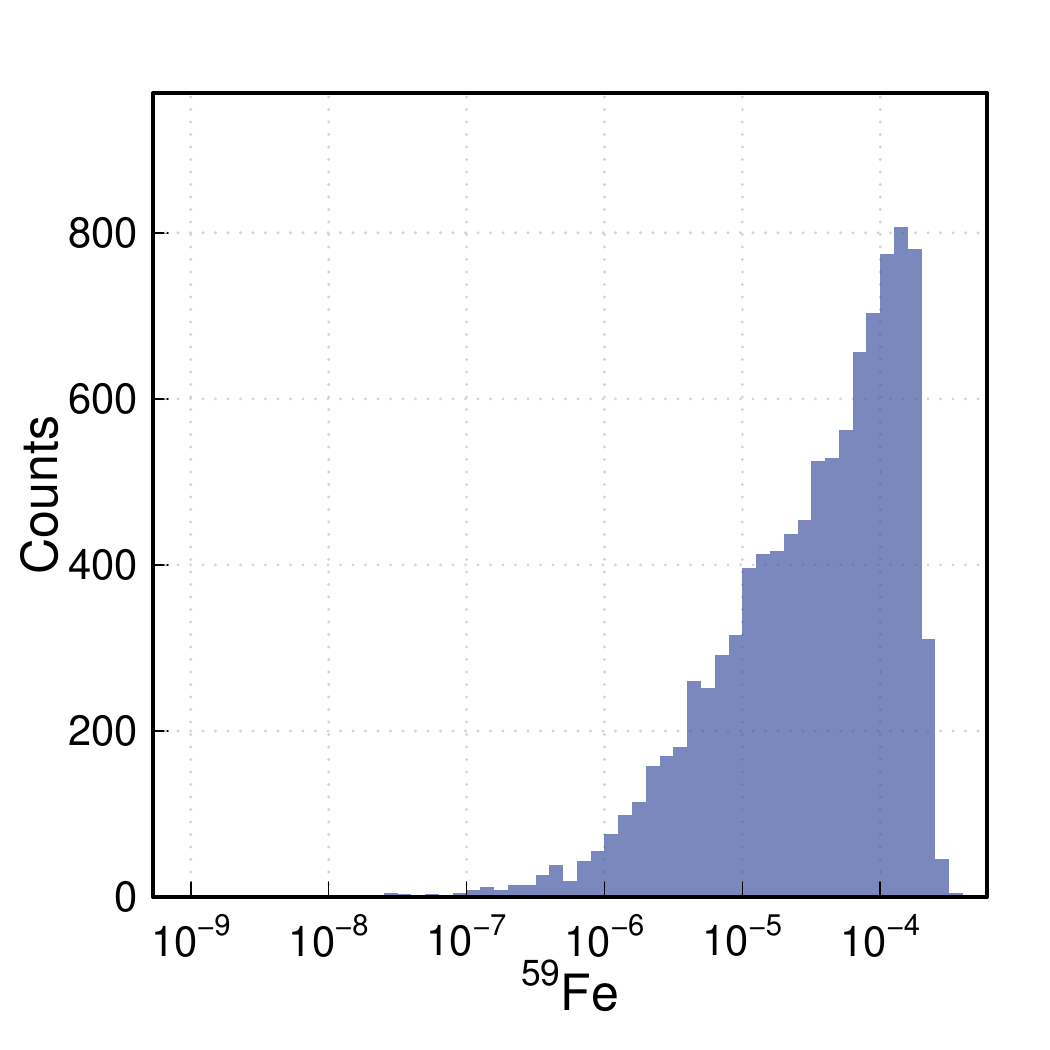}
    \includegraphics[width=0.33\linewidth]{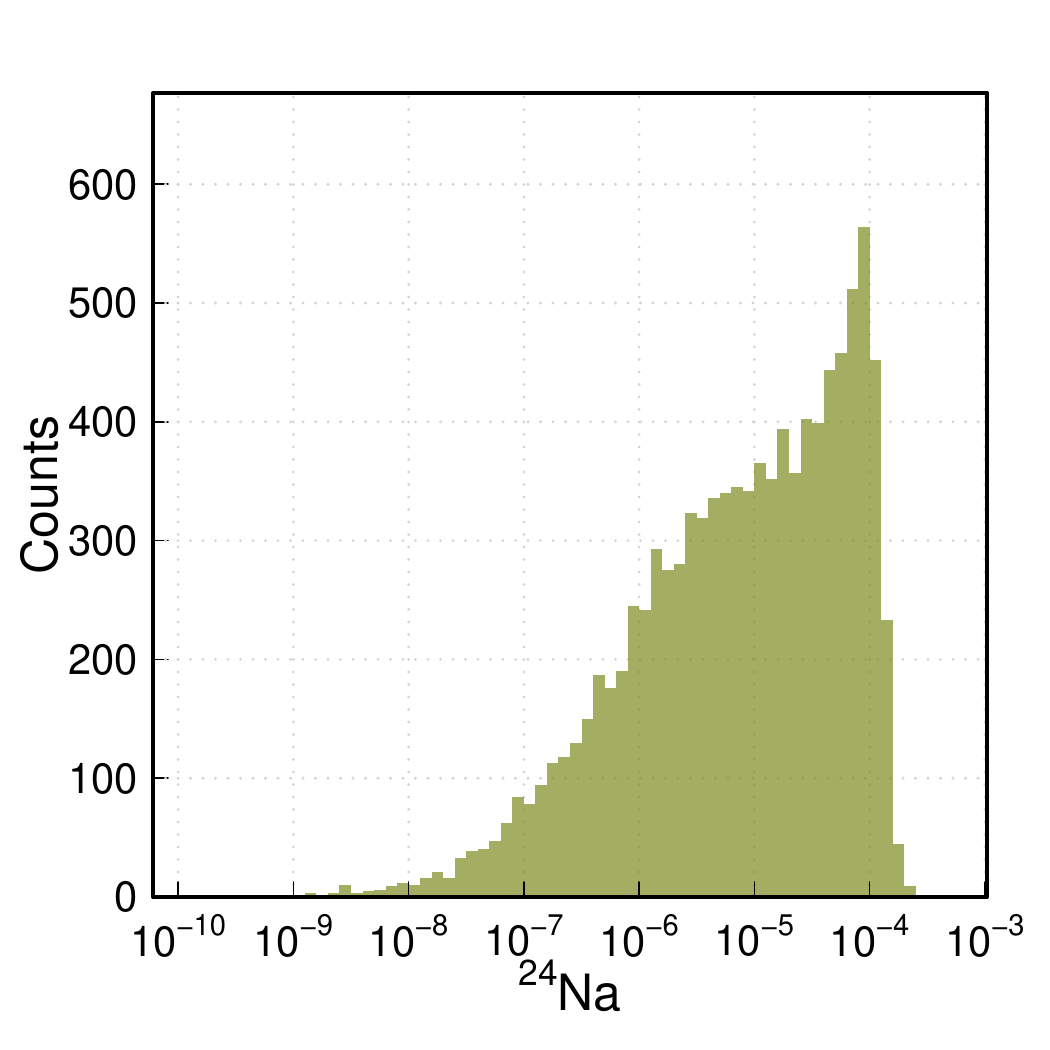}
    \caption{Variation in $^{51}$Cr, $^{59}$Fe, and $^{24}$Na production when nuclear reaction rates are varied within their default \texttt{STARLIB}
    uncertainties.  Compare to the $^{44}$Ti uncertainties in Figure~\ref{fig:44Ti-MC}.}
    \label{fig:MCshells}
\end{figure*}

There are three black crosses in Figure~\ref{fig:shell-nuc}. These indicate three trajectories that we selected as being representative of $^{24}$Na, $^{51}$Cr and $^{59}$Fe production, respectively. We computed nucleosynthesis for these three trajectories with multiple nuclear reaction network codes with their standard assumptions, for which the results were presented in Figure~\ref{fig:shell-network-comparison} in Section~\ref{sec:nuclear_networks}.
Furthermore, we used these trajectories to study the nuclear reaction rate uncertainties and, similarly to the $^{44}$Ti case, these can translate into considerable uncertainties in our nuclear yields.  Figure~\ref{fig:MCshells} shows the affect of nuclear-rate uncertainties on 3 shell-burning isotopes:  $^{51}$Cr, $^{59}$Fe and $^{24}$Na.  By considering the correlations between the yield of these nuclides with the reaction rate variations used in the Monte Carlo nucleosynthesis calculation, we can identify which reactions have uncertainties that most impact their production. For $^{51}$Cr, many reactions impact its yield, with $^{52}$Cr(p,n)$^{52}$Mg and $^{28}$Si($\alpha$,$\gamma$)$^{32}$S having the most impact. For $^{59}$Fe, uncertainties in the $^{58}$Fe(n,$\gamma$)$^{59}$Fe, $^{59}$Fe(n,$\gamma$)$^{60}$Fe, and $^{59}$Fe(p,n)$^{59}$Co reactions completely dominate the uncertainty in its yield. Finally, for the case of $^{24}$Na, uncertainties in the $^{12}$C+$^{12}$C $\rightarrow$ $^{23}$Na $+$ p and $^{24}$Na(p,n)$^{24}$Mg reactions are most responsible for the spread in this model.

Figure~\ref{fig:gammaray} includes the $\gamma$-ray signal from $^{59}$Fe produced in the helium shell of a supernova explosion.  Although this signal is dimmer than our $^{56}$Ni signal due to a much lower total production, for a Galactic supernova, $^{59}$Fe will be detected by modern $\gamma$-ray missions.

\section{Summary and Conclusions}
\label{sec:conclusions}

Nucleosynthetic yields, particularly through $\gamma$-ray observations, are an ideal probe of supernovae, their progenitors and their engines.  But to utilize these probes, we must minimize the uncertainties in the physics, physics implementation, and calculations.  Here we reviewed these uncertainties and our current understanding of the uncertainties.  The astrophysics uncertainties we discussed included:
\begin{itemize}
    \item Stellar Evolution:  mixing, burning, radiation transport, rotation, and mass loss
    \item Explosive Trajectories:  long-lived engines, shock deceleration, multi-dimensional and neutrino effects
\end{itemize}

We also included a discussion of the nuclear physics uncertainties focusing on the interplay between experimental and theoretical physics approach.  Right now, the nuclear rate uncertainties are as large as the astrophysical uncertainties, but a focused effort coupling experiments and improved theory is underway and can beat down these errors.  

Although most observations include a number of additional uncertainties, $\gamma$-ray observations are among the most direct, tying nuclear yields to either the supernova engine or its progenitor.  With considerable theory and experimental work, the existing uncertainties will be sufficiently reduced to allow these $\gamma$-ray observations to constrain the supernova engine and the physics behind it.  Considerable more effort is required to even determine whether UVOIR observations will be able to provide quantitative constraints on the central supernova engine.

%%\label{}

\section*{Acknowledgements}

The work for this paper was supported in part by the US Department of Energy through the Los Alamos National Laboratory. Los Alamos National Laboratory is operated by Triad National Security, LLC, for the National Nuclear Security Administration of U.S.\ Department of Energy (Contract No.\ 89233218CNA000001), the U.S. Department of Energy, Office of Science, Office of Nuclear Physics, under Award Number DE-SC0023128, and the National Science Foundation under Grant No. OISE-1927130.  We thank the Institute for Nuclear Theory at the University of Washington for its kind hospitality and stimulating research environment. This research was supported in part by the INT's U.S. Department of Energy grant No. DE-FG02- 00ER41132.

%% The Appendices part is started with the command \appendix;
%% appendix sections are then done as normal sections
\appendix

%\section{Appendix title 1}
%% \label{}

%% If you have bibdatabase file and want bibtex to generate the
%% bibitems, please use
%%
\bibliographystyle{elsarticle-harv} 
\bibliography{refs.bib}

%% else use the following coding to input the bibitems directly in the
%% TeX file.

%%\begin{thebibliography}{00}

%% \bibitem[Author(year)]{label}
%% For example:

%% \bibitem[Aladro et al.(2015)]{Aladro15} Aladro, R., Martín, S., Riquelme, D., et al. 2015, \aas, 579, A101

%%\end{thebibliography}

\end{document}